\documentclass[]{emulateapj}
\usepackage{graphicx}
\newcommand\mathnew{\mathsurround=0pt}
\def\simov#1#2{\lower .5pt\vbox{\baselineskip0pt \lineskip-.5pt
        \ialign{$\mathnew#1\hfil##\hfil$\crcr#2\crcr\sim\crcr}}}

\def\lesssim{\mathrel{\hbox{\rlap{\hbox{\lower4pt\hbox{$\sim$}}}\hbox{$<$}}}
}
\def\la{\mathrel{\hbox{\rlap{\hbox{\lower4pt\hbox{$\sim$}}}\hbox{$<$}}}}
\def\ga{\mathrel{\hbox{\rlap{\hbox{\lower4pt\hbox{$\sim$}}}\hbox{$>$}}}}

\shortauthors{Levine et al.}
\shorttitle{An Extended Periodicity Search}

\begin{document}

\title{An Extended and More Sensitive Search for Periodicities in RXTE/ASM X-ray Light Curves}
\author{Alan M. Levine\altaffilmark{1}, Hale V. Bradt\altaffilmark{1,2},
  Deepto Chakrabarty\altaffilmark{1,2}, Robin H. D. Corbet\altaffilmark{3}, and
  Robert J. Harris\altaffilmark{4}}
\altaffiltext{1} {Kavli Institute for Astrophysics and Space Research, MIT, Cambridge, MA 02139,
  USA; aml@space.mit.edu, hale@space.mit.edu, deepto@mit.edu}
\altaffiltext{2} {Physics Department, MIT, Cambridge, MA 02139,
  USA}
\altaffiltext{3}{University of Maryland, Baltimore County, and X-ray Astrophysics Laboratory, 
  Code 662, NASA/Goddard Space Flight Center, Greenbelt, MD 20771, USA;
  robin.corbet@nasa.gov}
\altaffiltext{4} {Harvard-Smithsonian Center for Astrophysics, 60 Garden St., Cambridge, MA 02138,
  USA; rjharris@cfa.harvard.edu}

\begin{abstract}

We present the results of a systematic search in $\sim14$ years of
{\it Rossi X-ray Timing Explorer} All-Sky Monitor data for evidence of
periodicities not reported by Wen et al. (2006).  Two variations of
the commonly used Fourier analysis search method have been employed to
achieve significant improvements in sensitivity. The use of these
methods and the accumulation of additional data have resulted in the
detection of the signatures of the orbital periods of eight low-mass
X-ray binary systems and of ten high-mass X-ray binaries not listed in
the tables of Wen et al.

\end{abstract}


\section{Introduction}

We have extended the search of \citet{wen06} for periodicities in the
intensities of X-ray sources using light curves produced by the {\it
Rossi X-ray Timing Explorer (RXTE)} All-Sky Monitor (ASM).
\citet{wen06} analyzed data accumulated during the first 8.5 years of
the mission and detected 41 periodicities and 5 possible periodicities
or quasiperiodicities.  Two of the periodicities detected in their
search likely represent neutron star spin and $\sim 6$ represent the
precession of tilted accretion disks or other phenomena that produce
superorbital periods.  The 33 other periodicities are believed to
represent evidence of the orbital motion of a close-binary star
system.  Most of the orbital periods were associated with high-mass
X-ray binaries (HMXBs) where the normal type star is either a
supergiant or a Be star.  Eight of the periodicities were associated
with the orbital periods of low-mass X-ray binaries (LMXBs), and one
was associated with a cataclysmic variable (CV).

The search of \citet{wen06} was one of the most comprehensive searches
for periodicities in the range of hours to years in the X-ray
intensities of the bright Galactic X-ray sources.  However, relatively
few new periods were found in the course of the search in spite of the
conventional wisdom that nearly all of the bright Galactic sources are
accreting binaries with periods in the range of sensitivity of the
search.  The orbital periods of many Galactic sources are still not
determined, and so searches for evidence of these presently unknown
orbital periods remain of interest.

The present search improves on the sensitivity achieved in the earlier
search through the analysis of light curves obtained over a longer
course of time, i.e. nearly 14 years as opposed to 8.5 years, and
through two major changes in the analysis techniques. We have also
analyzed subintervals of the mission-long data sets.  A preliminary
report of the results of this work was presented in \citet{lcatel06}.

\section{Data}\label{data}

The ASM consists of three Scanning Shadow Cameras (SSCs) mounted on a
rotating Drive Assembly \citep{asm96}.  Each SSC acts as a
coded-aperture camera and comprises a position-sensitive proportional
counter (PSPC) that detects X-ray photons and provides one-dimension
position information via the charge-division technique.  The entrance
window of each PSPC has a total geometric area of about 68 cm$^2$.  A
mask that is perforated with a pseudorandom pattern of open slits is
held 30 cm above the PSPC entrance window.  The mask open fraction is
close to 50\%.  The field of view of each SSC is $6\arcdeg \times
90\arcdeg$ (FWHM). The ASM observations have almost entirely been made
as series of 90-s exposures known as dwells.  In each dwell the aspect
of the assembly of SSCs is held fixed.  For each exposure, a
coded-aperture analysis of the data from each SSC yields intensity
estimates of the sources that are listed in a catalog (hereafter the
ASM catalog) and that fall within that SSC's field of view. The
intensity estimates are derived for each of four spectral bands which
nominally correspond to photon energy ranges of 1.5-3, 3-5, 5-12, and
1.5-12 keV and typically achieve a sensitivity of a few SSC counts
s$^{-1}$ (the Crab Nebula produces intensities of 27, 23, 25, and 75
SSC counts s$^{-1}$ in the 4 bands, respectively).  Any given source
has been viewed in a large number, e.g., 30,000 to 100,000, of the
90-s exposures over the $\sim14$ years of operation of the ASM.  Each
exposure generally yields intensity estimates for the four spectral
bands.

The present search has been carried out, except where otherwise noted,
using the ASM data acquired from the beginning of operations in early
1996 through 2010 January 28 (MJD 55224; Modified Julian Date = Julian
Date $-$ 2,400,000.5).

The coded-aperture analysis is applied to all ASM observations.  The
results are sorted source-by-source, filtered, and used to construct
light curve files for each of 571 potential sources for each of the
four energy bands.  The intensity measurements contained in these
files are the input data for the present periodicity search.

The instrument properties have changed over the 14 years of operation
in orbit because of at least three effects.  First, the gain of SSC 1
(of SSCs 1-3) increased at about 10\% per year for the first 13 years
or so because of a leak in the counter entrance window (the rate of
increase appears to have moderated in year 14).  This has resulted,
first, in progressive reductions in the effective sensitivity of SSC 1
in the 5-12 keV band and, recently, to some extent in the 3-5 keV band
because in the last few years the pulse heights of higher energy
events are often saturated and thus unusable.

The second of the effects is present because the one-dimensional
position-sensing capability of the proportional counters is based on
the charge-division technique and the resistivity of carbon-coated
quartz-fiber anodes.  The process of event detection in the counters
results in the redistribution or removal of the carbon. The
resistivity of each anode thus evolves over time generally becoming
larger and less uniform, and so the functions that relate charge ratio
to position have changed over the course of the mission.

Third, in the last few years there is often a high background of
events that have pulse heights that suggest that they are due to the
detection of Al K$\alpha$ photons.  We do not understand in detail why
this is happening but we speculate that it may be due to leakage of
ultraviolet radiation through the Al-coated Kapton thermal shields
that are immediately above the coded mask plates.  UV radiation would
produce photoelectrons at the counter body surfaces that happen to be,
by design of the instrument, at potentials of approximately $-1.8$
kV. The photoelectrons would be accelerated toward and would hit,
e.g., the aluminum collimators that are near the ground potential.
Some fraction of the Al K$\alpha$ photons that would be produced by
this process would then be detected.

Both the gain changes and charge ratio to position calibration changes
are accounted for in detail in the detector model used in our
coded-aperture analysis.  However, the calibrations are imperfect and
the derived source intensities are therefore somewhat affected by
systematic errors on both short and long time scales.  The coded
aperture analysis is also designed to eliminate the effects of
backgrounds that are smooth and slowly-varying with regard to position
in the detector; this is largely but not entirely effective.

\section{Analysis}
\label{tech}

Searches for periodicities where large numbers of measurements are
involved have generally proceeded by averaging the measured
intensities in equally-spaced time bins, Fourier transforming the
resulting time series, producing a power density spectrum by taking
the absolute square of the complex amplitude for each frequency, and
searching for peaks in the power spectrum.  Our search largely follows
this outline but we have implemented a few significant modifications
as described below.  In some searches for periodicities, a
Lomb-Scargle periodogram (\citealt{scargle82}; see also
\citealt{pressetal92}) is used in place of a Fourier transform, but,
when large numbers of measurements and time bins are being analyzed,
the differences between the two are not important except at extremely
low frequencies.  For the present analyses, the conventional discrete
Fourier transform is adequate so we do not make use of the
Lomb-Scargle variation.

We employ two unusual strategies with the goal of improving the search
sensitivity.  The first strategy involves the use of weights such as
the reciprocals of the variances in Fourier and other types of
analysis since the individual ASM measurements have a wide range of
associated uncertainties.  Since in the present instance the
amplitudes in the Fourier transforms may be regarded as statistical
averages, they may be better determined by the use of appropriate
weights.  This is discussed briefly in \citet{corbetal07} and at
greater length in \citet{corbmt07}. These two reports also discuss the
use of weights which involve the intrinsic source variability, but we
do not use such weights in the work reported herein.

The second strategy stems from the fact that the observations of the
source are obtained with a low duty cycle, i.e., the window function
is sparse but nevertheless highly structured.  The properties of the
window function, in combination with the presence of slow variations
of the source intensity, act to hinder the detection of variations on
short time scales.  The window function power density spectrum (PDS)
has substantial power at high frequencies, e.g., 1 cycle d$^{-1}$ and
1 cycle per spacecraft orbit ($\sim95$ minute period).  Since the data
may be regarded as the product of a (hypothetical) continuous set of
source intensity measurements with the window function, a Fourier
transform of the data is equivalent to the convolution of a transform
of a continuous set of intensity measurements with the window function
transform.  The high frequency structure in the window function
transform acts to spread power at low frequencies in the source
intensity to high frequencies in the calculated transform (or,
equivalently, the PDS).  This effectively raises the noise level at
high frequencies.  Thus, in most of our analyses, a smoothed version
of the light curve is subtracted from the binned data prior to the
Fourier analysis.

The details of the analysis of each ASM dwell-by-dwell light curve are
as follows. The time of each intensity measurement was adjusted to
approximately remove the effects of the motion of the Earth around the
Solar System barycenter and the adjusted times were used to assign the
measurements to equally-spaced time bins of duration 5 minutes
(0.0034722 d) or more.  Weights are used in averaging those
measurements that fall into a given time bin. The $n_i$ measurements
that fall into the time bin which is labelled by the index $i$ are
denoted with index $j$ that runs from 1 to $n_i$.  We denote the $j$th
measured source intensity in a time bin by $s_j$ and its associated
1-sigma uncertainty by $\sigma_j$.  The averaging in this stage uses
weights $w_j = 1/\sigma_j^{k_A}$ where $k_A$ is a selected weighting
index (see values in Table~\ref{tbl:mthds}).  The weighted-average
intensity in bin $i$ is then
\begin{equation}
d_i = \frac{\sum_{j=1}^{n_i} s_j/\sigma_j^{k_A}}{\sum_{j=1}^{n_i} 1/\sigma_j^{k_A}}
\end{equation}
For convenience we denote the
denominator of this expression by $W_i$ so that
\begin{equation}
W_i =  {\sum_{j=1}^{n_i} 1/\sigma_j^{k_A}}.
\end{equation}
where it should be understood that $W_i = 0$ for those time bins
that contain no measurements.  The $W_i$'s are used to weight the
time bins in the subsequent analysis.

\begin{deluxetable}{lccc}
\tablecolumns{4}
\tablewidth{0pt}
\tabletypesize{\footnotesize}
\tablecaption{Analysis Method Variations\label{tbl:mthds}}
\tablehead{
\colhead{Method} &
\colhead{Smoothing} &
\colhead{Weight} &
\colhead{Weight} \\
\colhead{Name} &
\colhead{Kernel} &
\colhead{Index $k_A$} &
\colhead{Index $k_B$}
}
\startdata
unwt & none\tablenotemark{a} & 2.0 & 0.0 \\
wt & none\tablenotemark{a} & 2.0 & 2.0 \\
ungs & Gaussian & 2.0 & 0.0 \\
wtgs & Gaussian & 2.0 & 2.0 \\
wtgs-v4 & Gaussian & 2.5 & 1.5 \\
wtgs-v8 & Gaussian & 2.0 & 3.0 \\
wtbx & box & 2.0 & 2.0
\enddata
\tablenotetext{a}{No subtraction of a smoothed version of the light curve
was done in this case.}
\end{deluxetable}

As stated above, a smoothed version of the light curve is, in many of
our analyses, subtracted from the binned data.  The smoothed version
is given by a time bin by time bin ratio in which the numerator is
computed by convolving a kernel function with a weighted version of
the binned light curve.  The denominator is computed by convolving the
same kernel function with a weighted version of the window function.
The value of the $i$th bin of the smoothed data is given by
\begin{equation}
D_i = \frac{(K \otimes (W \cdot d))_i}{(K \otimes W)_i}
\end{equation}
where $K$ represents the kernel function, $d$ and $W$ are the above
defined functions of the time bin index $i$, and ``$\otimes$'' denotes
convolution.  The kernel function $K$ may be either a Gaussian or a
box function.  In the case of the Gaussian kernel, the full width at
half maximum response is taken to be equal to the smoothing time
parameter.  The Gaussian is calculated out to $\pm 14$ standard
deviations from its center.  In the case of the box function, the box
width is taken to be twice the smoothing time parameter.  The
convolution is accomplished using Fourier transforms.  

We repeated the analyses using seven different values of the smoothing
time parameter (see Table~\ref{tbl:tmscls}) to optimize the
sensitivity in each of several different frequency ranges.  In each
case the smoothed light curve was subtracted from the unsmoothed light
curve, weights were applied, and the results were Fourier transformed.
In symbols, this part of the analysis produces a filtered light curve
described by:
\begin{equation}
F_i =  Y_i(d_i - D_i)
\end{equation}
where the weights $Y_i$ are given by
\begin{equation}
Y_i =  W_i^{k_B/k_A} = \left[{\sum_{j=1}^{n_i} 1/\sigma_j^{k_A}}\right]^{k_B/k_A}
\end{equation}
and $k_B$ is a second weighting index.  Weight index values are given
in Table~\ref{tbl:mthds}.
We also did the
analysis without subtracting a smoothed version of the light curve.

The unweighted average of the set of $F_i$'s which correspond to bins
with nonzero exposure is computed and then subtracted from each of the
$F_i$ values for those bins. The resultant array is extended with
zeroes by a factor of four and Fourier transformed.  The extension of
the array yields oversampling in the frequency domain.  The
oversampled transform is converted into a power density spectrum
which is normalized to have an average value of unity.

The normalized power spectra are compressed by saving the average and
maximum values for sets of contiguous frequency bins; the number of
contiguous bins depends on the frequency range and smoothing time
parameter (see Table~\ref{tbl:tmscls}).

\begin{figure*}[tbp]
\centering
\includegraphics[height=5.0in,angle=270.0]{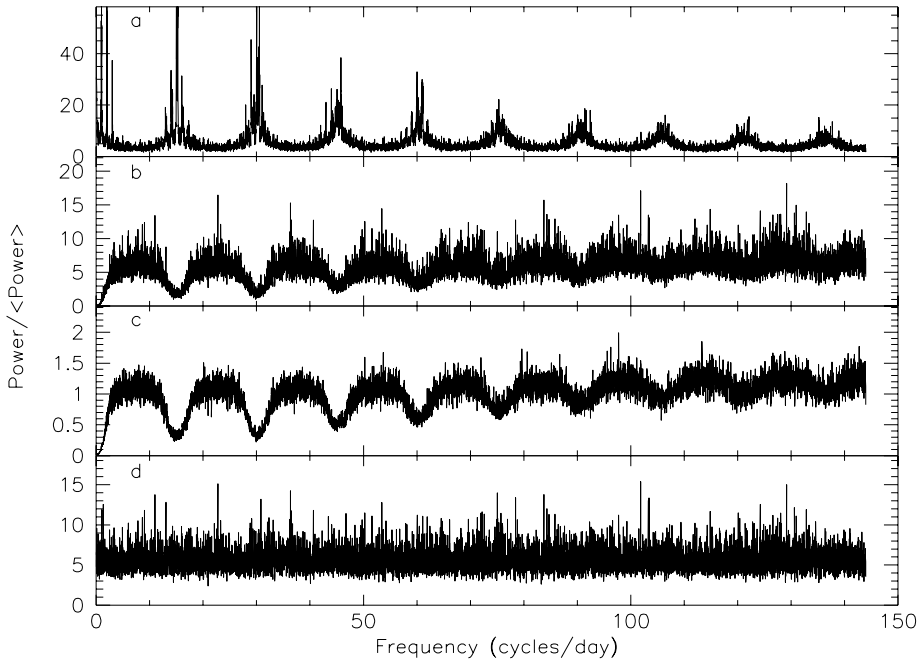}
\caption{Power density spectra (PDSs) based on the 1.5-12 keV ASM
light curve of LMC X-2. These PDSs were made using a smoothing time
scale of 0.3 days and have been compressed using the standard
parameters (Table~\ref{tbl:tmscls}); they illustrate the compression
and whitening procedure used in the search (see text).  The Nyquist
frequency is 144 d$^{-1}$. a) PDS of the window function, i.e., of the
weights used in the analysis of the data.  For this example the
weights are the reciprocals of the formally-estimated variances of the
binned data.  Those time bins of the window function which do not
contain any measurements are assigned weights of zero.  Prominent
peaks are seen at frequencies of 1 d$^{-1}$ and $\sim15$ d$^{-1}$,
i.e., the orbital frequency of the {\it RXTE} spacecraft, and
harmonics thereof. Some of the strong peaks extend substantially above
the upper boundary of the plot. b) The maximum values of each set of
300 contiguous bins of the light curve PDS. c) The average values of
each set of 300 contiguous bins of the light curve PDS. d) The
whitened maximum values of the PDS consisting of the maximum values
shown in (b) divided by a local (in frequency) background determined
from (c).  In this example, no power is seen that exceeds the
detection threshold.
\label{fig:whiten}}
\end{figure*}

We found that for the analyses where short smoothing time scales were
used, i.e., time scales of 0.3, 0.9, and 3.0 days, the compressed
power spectra have a non-white appearance.  The effect is strongest
for the shortest smoothing time scale, i.e., 0.3 days.  It is not
clear how to apply a threshold for detection of a periodicity in a
nonwhite spectrum.  Therefore we added a procedure (to the code that
performs the compression described above) to compute a ``background''
average power around each compressed frequency bin, and to then obtain
whitened maximum values.  Each whitened maximum value is then the
ratio of a maximum value in an unwhitened compressed spectrum to the
corresponding background average.  The resulting whitened spectra
generally appear to have little systematic structure (see
Fig.~\ref{fig:whiten}).

\begin{figure*}[tbp]
\centering
\includegraphics[height=5.0in,angle=270.0]{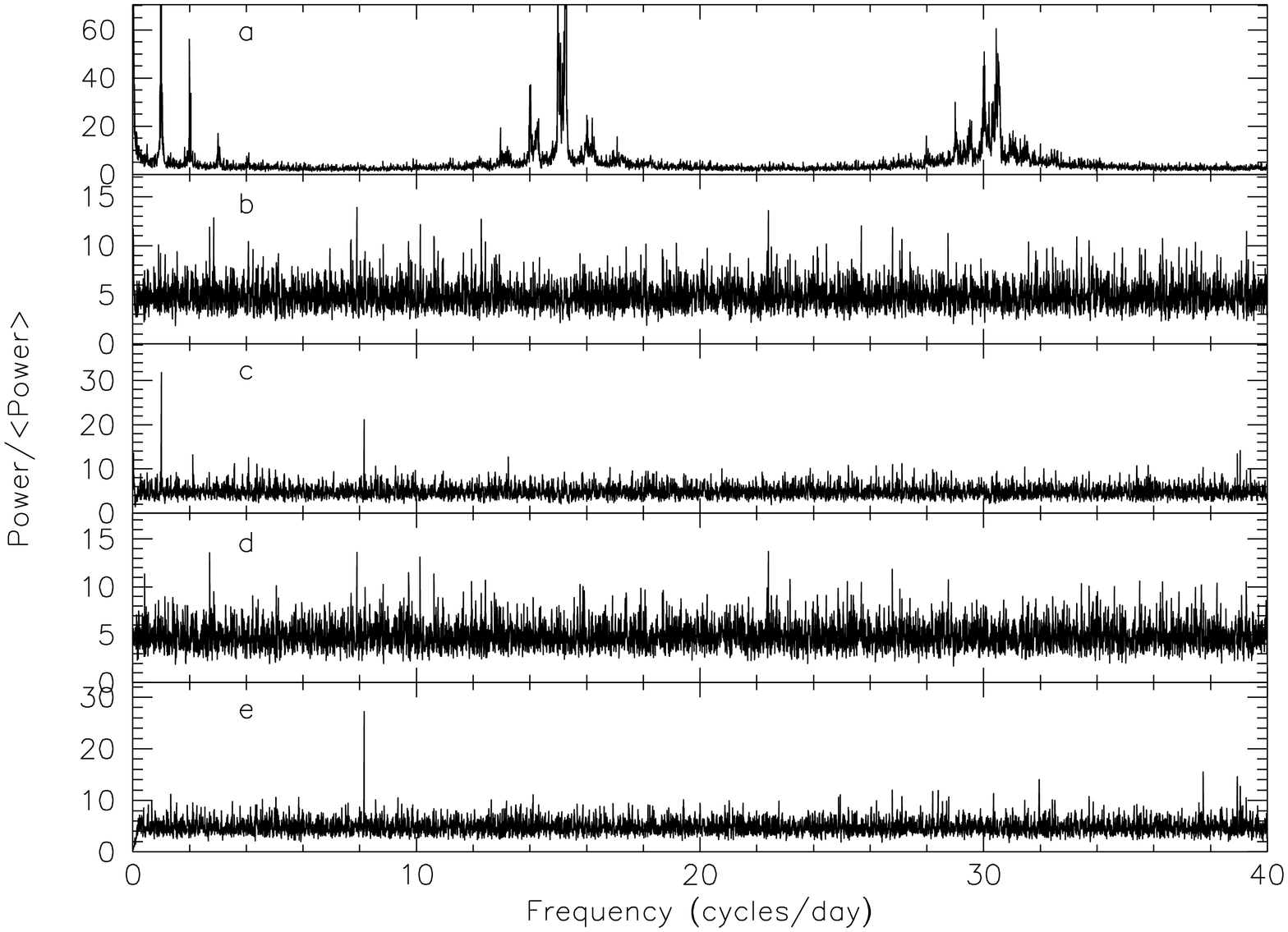}
\caption{Power density spectra (PDSs) based on the 1.5-12 keV ASM
light curve of X1323$-$619 that illustrate the typical differences in
the results obtained over a representative set of variations of the
analysis algorithm. The PDSs in panels b-e have been compressed using
the standard parameters for the 3.0 day smoothing time scale (see
Table~\ref{tbl:tmscls}); the whitened maximum powers are shown here.
The Nyquist frequency is 100 d$^{-1}$, but only frequencies up to 40
d$^{-1}$ are shown. a) PDS of the window function; see the caption for
Fig.~\ref{fig:whiten}. b) PDS computed with intensities that are
neither weighted nor filtered by subtraction of a smoothed version of
the light curve (``unwt'' entry in Table~\ref{tbl:mthds}). c) PDS
computed with intensities that are weighted but are not filtered
(``wt'' in Table~\ref{tbl:mthds}). d) PDS computed with intensities
that are not weighted but are filtered (``ungs'' in
Table~\ref{tbl:mthds}).  The light curve was smoothed using a 3.0 day
time scale for use in the filtering for the PDSs shown in this panel
and in panel e. e) PDS computed with intensities that are both
weighted and filtered (``wtgs'' in Table~\ref{tbl:mthds}). A signal is
apparent at the $\sim 8.15$ d$^{-1}$ orbital period of X1323$-$619
(See Table~\ref{tbl:detect}) in panel c but is seen with an even
higher amplitude in panel e. No peak stands out at this frequency
in either panel b or d.
\label{fig:pdscalc}}
\end{figure*}

Table~\ref{tbl:mthds} lists, at the highest level, the variations of
the analysis method that we have used.  We found that the results from
the methods where both weighting and filtering were applied (``wtgs'',
``wtgs-v4'', ``wtgs-v8'', and ``wtbx'') were almost always nearly
equal or superior to the results obtained with the other methods
(``unwt'', ``wt'', ``ungs'').  A typical, i.e., not the extreme best
case, of the differences is illustrated in
Figure~\ref{fig:pdscalc}. Therefore we did not search for significant
peaks in the power spectra obtained using the latter methods.  The
``wtgs'', ``wtgs-v4'', ``wtgs-v8'', and ``wtbx'' methods were all
applied to the 1.5-12 keV band light curves, but the results generally
produced power spectra that were quite similar in terms of both
detailed appearance and sensitivity. Therefore only the ``wtgs''
method was applied to the other three energy bands.  The results from
the four energy bands are, in some cases, quite different from each
other.  Table~\ref{tbl:tmscls} lists smoothing time scales and the
corresponding bin durations, parameters used for compression and
whitening, and parameters used in searching for significant peaks.

\begin{deluxetable*}{ccrrrcrr}
\tablecolumns{8}
\tablewidth{0pt}
\tabletypesize{\footnotesize}
\tablecaption{Filtering Timescales and Parameters\label{tbl:tmscls}}
\tablehead{
\colhead{Time Scale} &
\colhead{Bin Time} &
\colhead{N$_{\rm rebin}$} &
\colhead{N$_{\rm avg}$} &
\colhead{N$_{\rm some}$} &
\colhead{$\nu_{low}$} &
\colhead{$\nu_{hi}$} &
\colhead{Threshold} \\
\colhead{(days)} &
\colhead{(days)} &
 &
 &
 &
\colhead{(d$^{-1}$)} &
\colhead{(d$^{-1}$)} &
\colhead{Power}
}
\startdata
none & 2.0 & 1 & 300 & 3 & 0.0007 & 1.0 & 17.7 \\
500.0 & 2.0 & 1 & 300 & 3 & 0.004 & 1.0 & 17.7 \\
100.0 & 1.0 & 1 & 300 & 3 & 0.022 & 1.0 & 18.4 \\
30.0 & 0.15 & 10 & 60 & 1 & 0.1 & 4.0 & 20.3 \\
10.0 & 0.05 & 20 & 30 & 1 & 0.2 & 10.0 & 21.4 \\
3.0 & 0.005 & 150 & 12 & 1 & 0.66 & 100.0 & 23.7 \\
0.9 & 0.0034722222 & 300 & 6 & 1 & 2.2 & 144.0 & 24.1 \\
0.3 & 0.0034722222 & 300 & 6 & 1 & 6.6 & 144.0 & 24.1
\enddata
\end{deluxetable*}

The powers in a spectrum of Gaussian white noise are expected to be
well characterized by an exponential probability distribution. We make
the assumption that the powers in the whitened spectra computed as
described above similarly follow an exponential distribution.  In the
present case it is not clear how to precisely compute the number of
independent bins that we examined because 1) some of the power spectra
for a given source involve the same data and differ only on account of
slight variations in the analysis procedure, and 2) there is a
significant degree of dependence among the powers in an oversampled
power spectrum.  Although it is not fully justified on an a priori
basis, we simply estimate the number of independent bins $N_{ind}$ by
the product of the number of frequency bins that would have applied if
the Fourier spectrum had not been oversampled, the number of energy
bands analyzed (4; this neglects the non-independent bins among the
four slightly different analyses of the 1.5-12 keV band), and the
number of sources (571).  The neglect of the oversampling factor and
the multiple similar analyses of the sum band does not appear to have
led to significant underestimates of the detection thresholds, and,
thus, this procedure appears to be justified.

We have chosen a detection threshold $P_{thr}$ for previously unknown
periodicities to limit the expected number of false detections
$N_{exp}$ produced by statistical fluctuations (based on the
assumption that the powers are exponentially distributed) in the power
spectra calculated using a given smoothing time parameter.  Thus we
have
\begin{equation}
N_{exp} = N_{ind} e^{-P_{thr}}
\end{equation}
so that
\begin{equation}
P_{thr} = \log N_{ind} - \log N_{exp}.
\end{equation}
The values of $P_{thr}$ are given in Table~\ref{tbl:tmscls} for
$N_{exp} = 0.1$.

As can be seen in Figs.~\ref{fig:whiten} and \ref{fig:pdscalc}, the
window functions of the light curves often contain strong peaks at or
close to particular frequencies.  Even though the analysis procedure
tends to minimize the appearance of these frequencies in the whitened
power spectra, there are occurrences of peaks at these frequencies.
In the analyses done with filtering based on smoothing time scales of
10.0 days and under, we ignore all peaks in the frequency ranges
$1.00\pm0.04$, $2.00\pm0.04$, $15.15\pm0.20$, $30.3\pm0.2$,
$45.45\pm0.20$, and $60.6\pm0.2$ d$^{-1}$.  For the other analyses, we
ignore all peaks in the frequency ranges $0.00274\pm0.00025$,
$0.00548\pm0.00025$, $0.01096\pm0.00025$, $0.01644\pm0.00025$,
$0.01900\pm0.00035$, $0.01960\pm0.00025$, $0.02450\pm0.00025$, and
$0.02500\pm0.00025$ d$^{-1}$.

A substantial fraction of the 571 potential sources are
extragalactic targets such as active galactic nuclei, extended
Galactic targets such as supernova remnants, or X-ray binaries that
have been in quiescence since the beginning of the RXTE mission and
for which we have no expectation of finding detectable periodicities.
The power spectra of these sources act effectively as controls that
show the actual statistics of the power spectra and that bring
attention to periodicities that must be the product of statistical
fluctuations or artifacts.  Indeed, the search procedure described
here was tailored in part by examination of the results on the
``controls''.  Furthermore, since it would be difficult to understand
the reality of peaks that exceed the detection threshold in the power
spectra of these sources, the effective number of sources must be less
than 571. We have not adjusted the significance thresholds downward to
account for this effect.

Since we searched large numbers of bins, the thresholds for the
detection of previously unreported periodicities given in
Table~\ref{tbl:tmscls} are high.  When a periodicity has previously
been reported, it may be confirmed even if the power at or very close
to the previously reported frequency is below the threshold given in
the table.

Uncompressed power density spectra are useful for precise estimation
of the frequencies of the detected periodicities.  Since the
uncompressed spectra were not saved in the original analysis for
smoothing time scales less than or equal to 30.0 days, the analyses
that produced the most significant detection of each periodicity were
repeated, if necessary, and the resulting uncompressed spectra were
saved.  For each periodicity a small frequency range around the major
peak was fit with a Gaussian to obtain estimates of the central
frequency, its uncertainty, and the peak width.  The uncertainty in
the frequency may be estimated based on the result of Horne and
Baliunas (1986) that
\begin{equation}
\delta f_1 = \frac{3}{8}\frac{1}{T \sqrt{P_r}}
\end{equation}
where $T$ is the length of the time interval covered by the light
curve and $P_r$ is the peak power in a normalized PDS.  We also use
the simple estimate
\begin{equation}
\delta f_2 = \frac{1}{2T}
\end{equation}
which, in most cases, should be a conservative upper limit to the
uncertainty (but, if the periodicity persisted for only a fraction of
the time covered by the light curve, the uncertainty could be larger
than this value).  In the estimates of frequencies and periods that we
present below, the uncertainty derived using eq. (8) is given first
and the uncertainty derived using eq. (9) is given second in square
brackets.

\section{Results}\label{results}

In the present search we find clear evidence of the periodicities in
GRO J1008$-$57 and SS 433 that have been previously detected in the
ASM data but had not been detected in the analysis reported by
\citet{wen06}.  \citet{wen06} reported the detection of the 9.56-day
period of 4U 2206+543, but did not find evidence of variation with a
period of $\sim 19.1$ days.  Such variation was first reported by
\citet{corbmt07}.  Since we now find clear evidence of variation at
the $\sim 19.1$-d period, we report its detection herein.

In the previous section, we noted that periodicities at previously
identified frequencies can be found with better sensitivity than
periodicities at undetermined frequencies. Therefore we
examined the power spectra for peaks at the frequencies corresponding
to the periods listed in the catalogs of \citet{igrcat07} and
\citet{hmxb06,lmxb07} and thereby identified evidence of the presence
of a number of periodicities that were not found by \citet{wen06}.

Table~\ref{tbl:detect} lists the periodicities that emerge from the
present search that are not listed in Tables 1-3 of \citet{wen06}.
Only one of these was, when first detected in the course of this
search, previously unreported on the basis of optical or other X-ray
observations, namely the 18.545-day periodicity in IGR J18483$-$0311.
The photon energy band, filter time scale, and method that yielded the
most significant detection are given for each detected periodicity.
The estimated frequencies and associated uncertainties are listed in
Table~\ref{tbl:detect} with the exception of the very low frequencies
of the sources GRO J1008$-$57 and 2S 1845$-$024.  The frequencies or,
equivalently, periods of these latter sources are better obtained from
epoch-folding analyses and even more precisely from pulse timing
studies.

We give in Table~\ref{tbl:ampl} average source intensities, peak
powers seen in the whitened and renormalized power spectra, and
estimated average amplitudes of the detected periodic signals.  The
source names, photon energy bands, and periods are taken from
Table~\ref{tbl:detect}.  

The estimated average amplitudes of the detected periodic signals were
obtained by superposing sine waves onto the measured intensities in
the appropriate light curves and redoing the complete analysis using
the appropriate method, time scale, and energy band.  The sine wave
frequencies were chosen to yield peaks in the power spectrum close to,
but not on top of, the frequencies of the detected signals.  The sine
wave amplitudes were chosen to give peak powers close to or above,
within a factor of two, the peak powers of the detected signals.
The results were then interpolated to estimate the amplitudes that
would give rise to the peak powers of the detected signals. The
uncertainty in the amplitude for a given source was estimated by the
rms dispersion in the amplitudes obtained when sine waves were
injected at four to ten different frequencies.

In the last column of Table~\ref{tbl:ampl} we give the ratio of the
inferred average signal amplitude (A) to the weighted average source
intensity (I).  This is an indicator of the fractional variation of
the source intensity at the orbital frequency. It does not include the
effects of harmonics which are detected in some of the power density
spectra and should, therefore, be used only as a general indicator of
the degree of modulation.

\begin{deluxetable*}{llcccccc}
\tablecolumns{8}
\tablewidth{0pt}
\tabletypesize{\scriptsize}
\tablecaption{Additional Periodicity Detections\label{tbl:detect}}
\tablehead{
\colhead{Source Name\tablenotemark{a}} &
\colhead{ASM Catalog}      &
\colhead{Band} &
\colhead{Filter Time\tablenotemark{b}} &
\colhead{Method\tablenotemark{c}} &
\colhead{Frequency\tablenotemark{d}} &
\colhead{Period\tablenotemark{d}} &
\colhead{Epoch\tablenotemark{e}} 
 \\
  &
\colhead{Entry} &
\colhead{(keV)} &
\colhead{(days)} &
\colhead{} &
\colhead{(cycles d$^{-1}$)} &
\colhead{} &
\colhead{(MJD)}
 }
\startdata



\cutinhead{LMXB orbital periods}

2S\,0921$-$630 & x0921$-$630 & 1.5-12 & 30.0 & wtbx & 0.110994 (13; 100) &
 216.23 (3; 19) h & 52738.13 (13) \\
4U\,1254$-$69 & x1254$-$690 & 1.5-12 & 3.0 & wtgs-v4 & 6.101690 (15; 100) &
 3.933337 (10; 64) h & 52733.0546 (31) \\
4U\,1323$-$62 & x1323$-$619 & 1.5-12 & 3.0 & wtgs & 8.157930 (15; 100) &
 2.941923 (5; 36) h & 52733.2395 (18) \\
4U\,1636$-$536 & x1636$-$536 & 1.5-12 & 0.9 & wtgs & 6.327232 (14; 100) &
 3.793128 (8; 60) h & 52733.0343 (18) \\
4U\,1728$-$16\tablenotemark{j} & gx9+9 & 1.5-12 & 0.9 & wtgs & 5.723207 (7; 100) &
 4.193453 (5; 73) h & 52733.152 (1) \\
4U\,1746$-$37 & x1746$-$370 & 1.5-12 & 3.0 & wtgs-v8 & 4.648202 (15; 100) &
 5.163287 (18; 111) h & 52733.156 (3) \\
GRS\,1758$-$258 & grs1758$-$258 & 3-5 & 500.0 & wtgs & 0.052706 (19;100) & 
 18.973 (7; 36) d & 52748.27 (23) \\
4U\,1820$-$30 & x1820$-$303 & 1.5-12 & 0.9 & wtgs & 126.129380 (13; 100) &
 0.19028081 (2; 15) h & 52733.00590 (7) \\

\cutinhead{HMXB orbital periods}

A\,0535+26 & x0535+26 & 1.5-12 & 500.0 & wtgs & 0.00906\tablenotemark{f} & 
 110.4\tablenotemark{f} & 52852.59 (27) \\
LMC X-1 & lmcx1 & 1.5-12 & 100.0 & wtbx & 0.255821 (14; 100) &
 3.90898 (21; 153) d & 52733.952 (45) \\
GRO\,J1008$-$57 & groj1008$-$57 & 3-5 & 500.0 & wtgs &  &
  \tablenotemark{g} &  \\
2S\,1417$-$62 & x1417$-$624 & 5-12 & 100.0 & wtgs & 0.023611 (45; 100) &
 42.35 (8; 18) d & 52770.04 (91) \\
IGR\,J16320$-$4751 & igrj16320$-$4751 & 5-12 & 100.0 & wtgs & 0.111234 (11; 100) &
 8.9901 (9; 81) d & 52742.79 (12) \\
IGR\,J16418$-$4532 & igrj16418$-$4532 & 5-12 & 30.0 & wtgs & 0.267461 (20; 100) &
 3.73886 (28; 140) d & 52735.84 (7) \\
EXO\,1722$-$363 & exo1722$-$363 & 5-12 & 30.0 & wtgs & 0.102655 (19; 100) &
 9.7414 (18; 95) d & 52738.25 (19) \\
IGR\,J18027$-$2016 & igrj18027$-$2017 & 5-12 & 30.0 & wtgs & 0.218881 (16;100) &
 4.5687 (3;21) & 52736.36 (42) \\
2S\,1845$-$024 & x1845$-$024 & 5-12 & 500.0 & wtgs &  &
  \tablenotemark{h} &  \\
IGR\,J18483$-$0311 & igrj18483$-$0311 & 5-12 & 100.0 & wtgs &  0.053923 (9; 100) &
 18.545 (3; 34) d & 52754.37 (10) \\
3A\,1909+048\tablenotemark{k} & ss433 & 5-12 & 100.0 & wtgs & 0.076452 (20; 100) &
 13.080 (3; 17) d & 52745.19 (18) \\
4U\,2206+543 & x2206+543 & 1.5-12 & 100.0 & wtbx & 0.104620 (13; 100)\tablenotemark{i}  &
 9.5584 (12; 91)\tablenotemark{i} d & 52740.97 (6) \\

\enddata

\tablenotetext{a}{The source name is generally that listed first by
\citet{lmxb07} or \citet{hmxb06}.}

\tablenotetext{b}{See text and column 1 of Table~\ref{tbl:tmscls}.}

\tablenotetext{c}{See text and Table~\ref{tbl:mthds}.}

\tablenotetext{d}{Estimated from the peak centroid.  The uncertainties
are shown in parentheses in units of the least significant digit and
were obtained from eqs. (8) and (9).}

\tablenotetext{e}{Epoch of X-ray minimum or maximum (see phase zero in
each light curve shown in Figs.~\ref{fig:fold1} and \ref{fig:fold2}).}

\tablenotetext{f}{The ASM power spectrum indicates the periodicity is
present but is not useful for precisely estimating the orbital
frequency and period. The values given are the best estimate from the
ASM spectrum.  The orbital period is estimated to be $110.3 \pm 0.3$
days via pulse timing (see text).  The corresponding frequency is
$0.009066 \pm 0.000025$ d$^{-1}$.}

\tablenotetext{g}{A peak power of 12 is seen in the second harmonic
(first overtone).  See Fig.~\ref{fig:pdsj1008}.}

\tablenotetext{h}{A peak power of 11 is seen in the second harmonic
(first overtone).  See Fig.~\ref{fig:pdsj1008}.}

\tablenotetext{i}{The frequency and period given are those estimated
from the strongest peak in the power spectrum.}

\tablenotetext{j}{4U\,1728$-$16 $=$ GX 9+9}

\tablenotetext{k}{3A\,1909+048 $=$ SS\,433}

\end{deluxetable*}

\begin{deluxetable*}{lccccccc}
\tablecolumns{8}
\tablewidth{0pt}
\tabletypesize{\footnotesize}
\tablecaption{Characteristics of Detected Periodicities\label{tbl:ampl}}
\tablehead{
\colhead{} &
\colhead{} &
\colhead{Wtd. Avg.} &
\colhead{Unwtd. Avg.} &
\colhead{Approx.} &
\colhead{} &
\colhead{} &
\colhead{}
\\
\colhead{Source Name} &
\colhead{Band} &
\colhead{Intensity\tablenotemark{a}} &
\colhead{Intensity\tablenotemark{b}} &
\colhead{Period} &
\colhead{Power\tablenotemark{c}} &
\colhead{Amplitude\tablenotemark{d}} &
\colhead{A/I\tablenotemark{e}}
 \\
\colhead{} &
\colhead{(keV)} &
\colhead{(cts s$^{-1}$)} &
\colhead{(cts s$^{-1}$)} &
\colhead{} &
\colhead{} &
\colhead{(cts s$^{-1}$)} &
\colhead{}
 }
\startdata

\cutinhead{LMXB orbital periods}

2S\,0921$-$630 & 1.5-12 & 0.239\,(4) & 0.084\,(6) & 216.2 h & 23 & 0.052 & 0.22 \\
4U\,1254$-$69 & 1.5-12 & 2.455\,(4) & 2.291\,(6) & 3.933 h & 22 & 0.052 & 0.021 \\
4U\,1323$-$62 & 1.5-12 & 0.575\,(4) & 0.449\,(6) & 2.942 h & 27 & 0.055 & 0.096 \\
4U\,1636$-$536 & 1.5-12 & 9.069\,(6) & 9.339\,(9) & 3.793 h & 22 & 0.087 & 0.0096 \\
4U\,1728$-$16 (GX\,9+9) & 1.5-12 & 18.879\,(8) & 19.01\,(1) & 4.193 h & 123 & 0.283 & 0.015 \\
4U\,1746$-$37 & 1.5-12 & 2.275\,(6) & 2.06\,(1) &  5.163 h & 30 & 0.094 & 0.041 \\
GRS\,1758$-$258 & 3-5 & 0.485\,(5) & 0.391\,(7) & 18.97 d & 13 & 0.054 & 0.11 \\
4U\,1820$-$30 & 1.5-12 & 20.347\,(8) & 20.63\,(1) & 0.1903 h & 30 & 0.12 & 0.0059 \\

\cutinhead{HMXB orbital periods}

A\,0535+26 & 1.5-12 & 0.275\,(4) &  0.550\,(6) & 110.4\tablenotemark{f} & 11 & 0.21 & 0.78 \\
LMC X-1 & 1.5-12 & 1.524\,(4) & 1.369\,(5) & 3.909 d & 29 & 0.064 & 0.042 \\
GRO\,J1008$-$57 & 3-5 & 0.036\,(2) & 0.006\,(3) & & 10 \tablenotemark{g} & & \\
2S\,1417$-$62 & 5-12 & 0.096\,(3) & 0.035\,(6) & 42.4 d & 12 & 0.035 & 0.36 \\
IGR\,J16320$-$4751 & 5-12 & 0.224\,(4) & 0.119\,(7) & 8.99 d & 34 & 0.063 & 0.28 \\
IGR\,J16418$-$4532 & 5-12 & 0.165\,(5) & 0.060\,(9) & 3.739 d & 17 & 0.047 & 0.28 \\
EXO\,1722$-$363 & 5-12 & 0.132\,(4) & 0.029\,(7) & 9.741 d & 19 & 0.051 & 0.39 \\
IGR\,J18027$-$2016 & 5-12 & 0.189\,(7) & 0.03\,(1) & 4.569 d & 16 & 0.069 & 0.37 \\
2S\,1845$-$024 & 5-12 & 0.069\,(3) & -0.065\,(7) &  & 5 \tablenotemark{h} & \\
IGR\,J18483$-$0311 & 5-12 & 0.115\,(3) &  -0.021\,(7) & 18.54 d & 60 & 0.099 & 0.86 \\
3A\,1909+048 (SS\,433) & 5-12 & 0.256\,(3) & 0.189\,(6) & 13.08 d & 20 & 0.037 & 0.14 \\
4U\,2206+543 & 1.5-12 & 0.376\,(3) & 0.252\,(5) & 9.558 d 
  & 17\tablenotemark{i} & 0.08 & 0.21 \\

\enddata

\tablenotetext{a}{Weighted average of the entire ASM light curve
through MJD 55196 for the energy band given in column 2.  A formal
error based on the formal errors of the individual measurements is
shown. For reference, the average intensities of the Crab Nebula are
75.503\,(9), 26.825\,(5), 23.290\,(4), and 25.370\,(5) cts s$^{-1}$ in
the 1.5-12, 1.5-3, 3-5, and 5-12 keV bands, respectively.  Note that
the units of cts s$^{-1}$ are used as the general ASM proxy for source
intensity and not for observed count rates, i.e., they give intensity
in terms of the count rates due to the given source that would have
been seen had SSC 1 been in the same physical condition as it was in
March 1996 and had the source been in the center of its field of
view.}

\tablenotetext{b}{Unweighted average of the entire ASM light curve
through MJD 55196 for the energy band given in column 2.  A formal
error based on the formal errors of the individual measurements is
shown.  The unweighted average intensity is given for comparison with
the weighted average; the difference is a rough indicator of the
actual uncertainty in the average intensity.}

\tablenotetext{c}{Approximate peak power in the whitened and
renormalized power density spectrum (see text) at the frequency given
in Table~\ref{tbl:detect}.}

\tablenotetext{d}{Amplitude of a constant-amplitude sine wave that,
when superposed onto the actual data, yields a peak power close to
that given in column 6 (see text).}

\tablenotetext{e}{Ratio of the inferred average signal amplitude (A;
from column 7) to the weighted average source intensity (I; from
column 3).}

\tablenotetext{f, g, h}{\,See notes (e), (f), and (g) to
Table~\ref{tbl:detect}.}

\tablenotetext{i}{Power of the peak near half the frequency given in
Table~\ref{tbl:detect} or twice the period given in column 5.}

\end{deluxetable*}

\begin{figure*}[tbp]
\includegraphics[width=8.0in,angle=0.0]{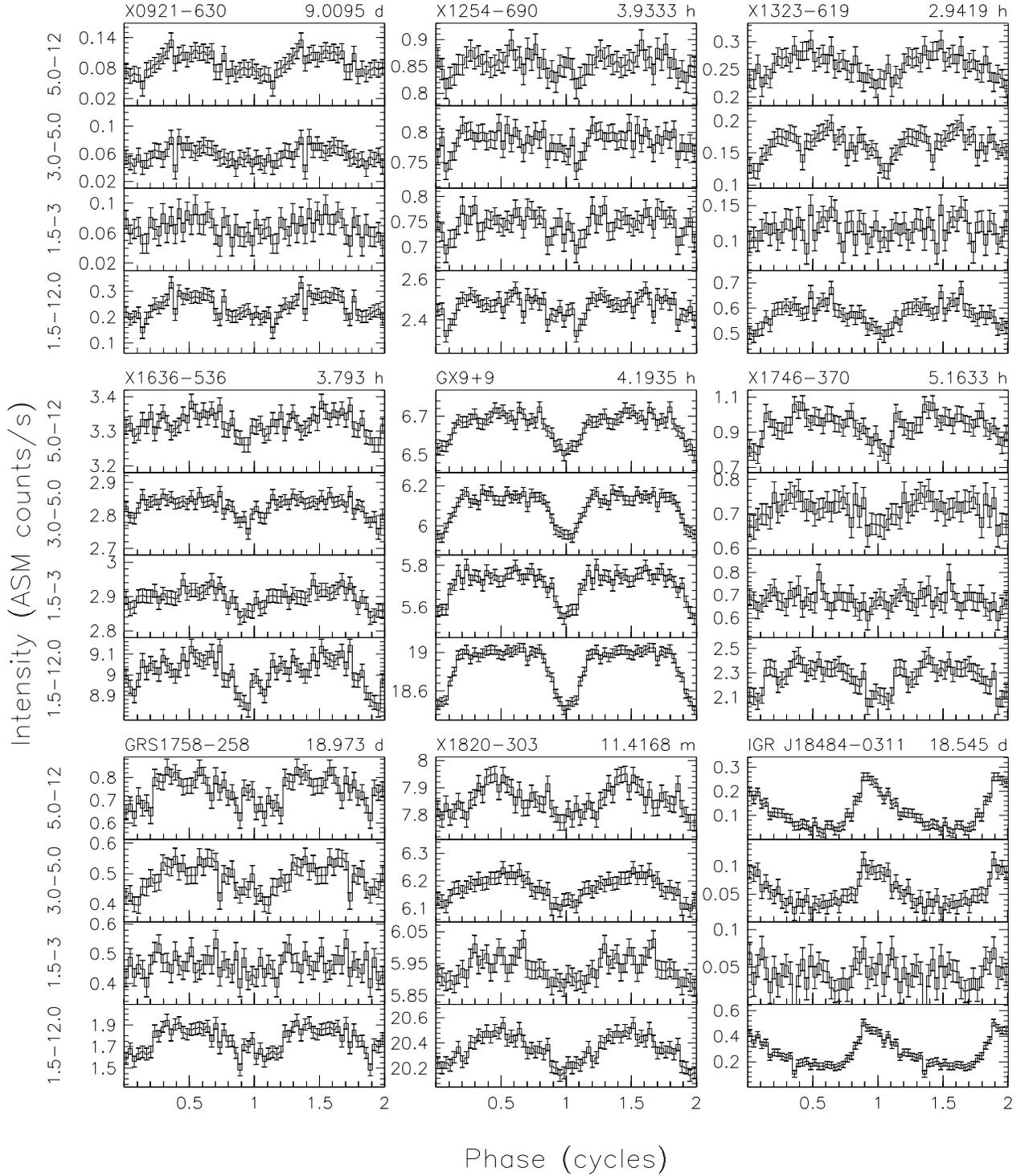}
\caption{ASM light curves of 9 sources folded at the orbital periods
given in Table~\ref{tbl:detect}.  See the text for information on the
details of the folding algorithm and the error bars shown here. For
each source the folded light curves are shown for two full cycles for
each of four energy bands.
\label{fig:fold1}}
\end{figure*}

\begin{figure*}[tbp]
\includegraphics[width=8.0in,angle=0.0]{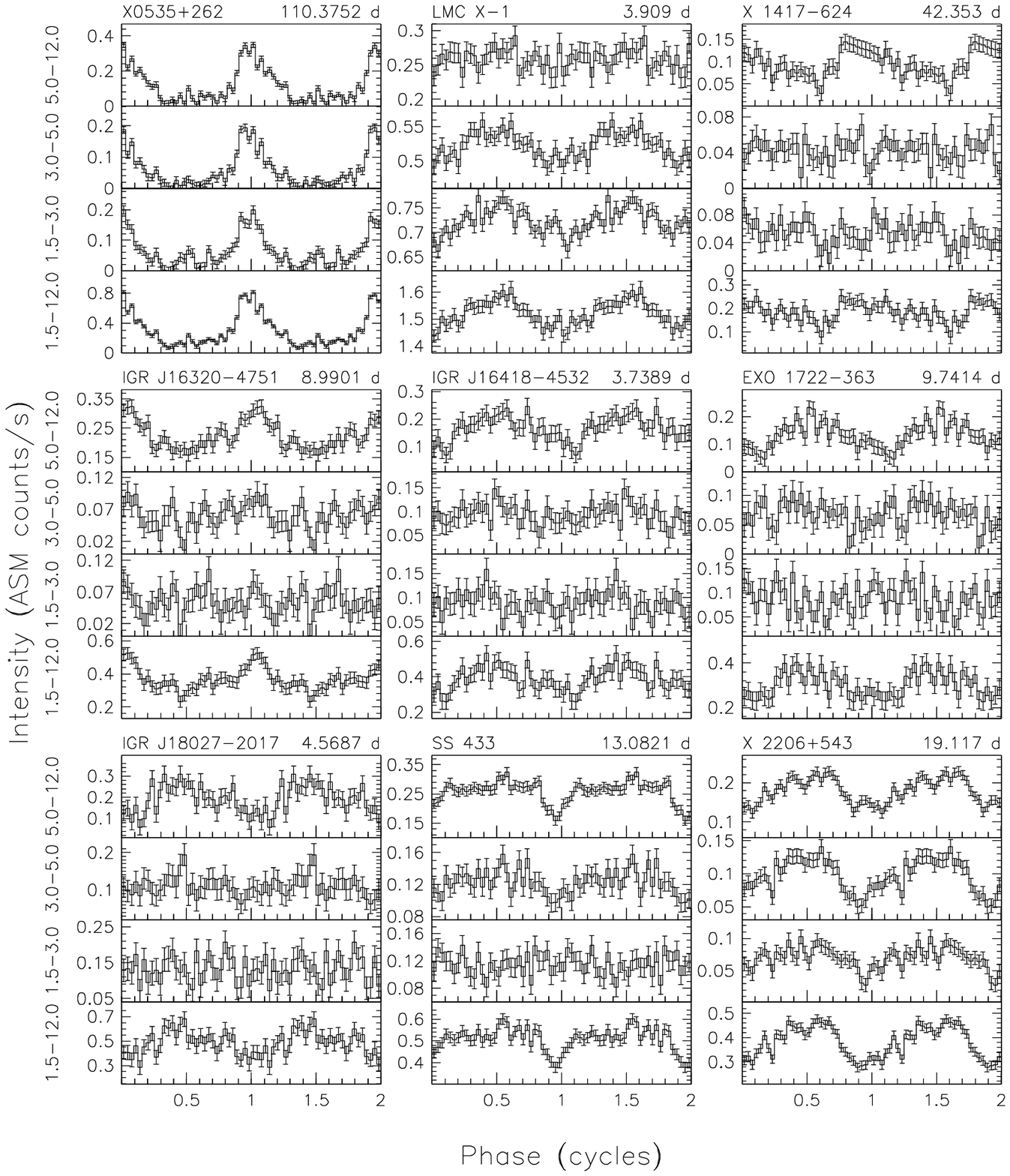}
\caption{ASM light curves of 9 sources folded at the orbital periods
given in Table~\ref{tbl:detect}.
\label{fig:fold2}}
\end{figure*}

The ASM light curves of the sources for which we report detections of
periodicities in Table~\ref{tbl:detect} were folded at periods close
to, i.e., within a small fraction of the quoted 1\,$\sigma$ errors,
those listed in that table.  Weights were used in averaging the
intensities in each phase bin.  A smoothed version of the light curve
was subtracted from each light curve before it was folded. The time
scales used for the smoothing are listed in
Table~\ref{tbl:detect}. After the folding was accomplished, the
overall average intensity was then added to the folded light curve to
show the proper mean (the mean was removed when the smoothed version
was subtracted from the light curve).  The folded light curves are
shown in Figures~\ref{fig:fold1} and \ref{fig:fold2}.

The uncertainties shown in the plots of the folded light curves were
determined from propagation of the 1\,$\sigma$ uncertainties associated
with each measurement in a light curve assuming the measurements are
statistically independent.  This neglects the effects of random source
variability in at least two respects.  First, the effects of source
variability are not included in the formal uncertainties of the
individual measurements.  Second, when variability is present it
almost always involves correlations from measurement to measurement.

The epochs of phase zero used in plotting the folded light curves are
listed in Table~\ref{tbl:detect}.  These epochs were determined by
fitting each folded 1.5-12 keV band light curve with a combination of
constant, sine, and cosine functions of phase.  The epoch is then
obtained from the phase of minimum or maximum intensity of the fitted
function. Actual minima or maxima may deviate somewhat from these
epochs; the differences may be seen in the folded light curves.

\subsection{Additional Detections of Low-Mass X-ray Binaries}

In this section we briefly discuss our results for each of the
periodicities of the LMXBs listed in Table~\ref{tbl:detect}.  The
results in Table~\ref{tbl:detect} show that, with only one exception,
the most significant detections of these systems were made in the
overall 1.5-12 keV energy band.  

\begin{figure}[tbp]
\includegraphics[height=3.5in,angle=0.0]{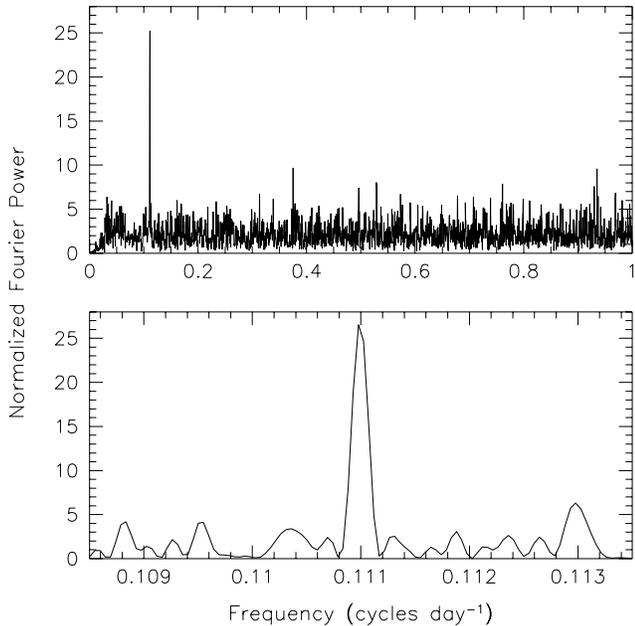}
\caption{(top) Portion of a compressed power density spectrum made
from the 1.5-12 keV light curve of 2S\,0921$-$630.  The ``wtbx'' method
and a 30.0 day smoothing time scale were employed in producing the
PDS. (bottom) A small region of the corresponding uncompressed PDS.
\label{fig:pds0921}}
\end{figure}

{\bf 2S\,0921$-$630} is a weak X-ray source in a low-mass system with
a relatively long orbital period.  Photometry of the bright optical
counterpart by \citet{ci81} and \citet{brand81} suggested that the
observed variability could be periodic, but did not yield correct
estimates of the actual period. X-ray observations with the {\it
EXOSAT} satellite showed intensity changes, interpreted as partial
eclipses, that were key to obtaining the first relatively precise
period estimate \citep{mason87}.  Recent optical observations have
also yielded relatively precise estimates of the orbital period;
\citet{shah04} obtained $P = 9.0035 \pm 0.0029$ d and \citet{jonk05}
obtained $P = 9.006 \pm 0.007$ d. The orbital period is detected in
the present study; this is shown in the power spectrum plotted in
Fig.~\ref{fig:pds0921}.  We derive an estimate of the orbital period
of $P = 9.009 \pm 0.001\,[\pm 0.008]$ d (see Table~\ref{tbl:detect})
that is consistent with these recently reported values.
\citet{shahwat07} and \citet{stejon07} report further work on the
binary parameters including the component masses. The folded ASM light
curves show that the periodic modulation has an amplitude of
$\sim$30\% relative to the average source strength.  However, the
average source intensity could be affected by systematic errors in the
baseline level of order 0.1 SSC counts s$^{-1}$ (1.5-12 keV band).
Nonetheless, the modulation fraction must be rather large, and the
periodicity may be easily detectable by other instruments.

\begin{figure}[tbp]
\includegraphics[height=3.5in,angle=0.0]{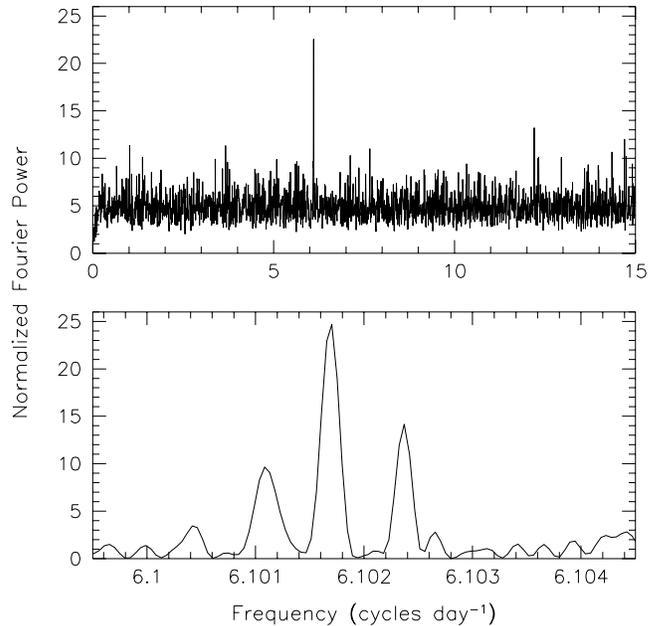}
\caption{(top) Portion of a compressed power density spectrum made
from the 1.5-12 keV light curve of 4U\,1254$-$69.  The ``wtgs-v4'' method
and a 3.0 day smoothing time scale were employed in producing the
PDS. (bottom) A small region of the corresponding uncompressed PDS.
\label{fig:pds1254}}
\end{figure}

{\bf 4U\,1254$-$69} is a so-called dipping LMXB.  Its $\sim 3.9$ hr
orbital period was revealed in both X-ray and optical observations
\citep{courv86,motch87}. \citet{diaztr09} report the results of recent
observations of this source with the {\it XMM-Newton} and {\it
INTEGRAL} satellites.  They also discuss the detection of the 3.9-hr
periodicity in ASM data, present a folded light curve, and use the ASM
data to obtain an estimate of the orbital period of $P = 0.16388875
\pm 0.00000017$ d ($= 3.933330 \pm 0.000004$ h).  We also detect the
orbital period in the ASM light curves (see Fig.~\ref{fig:pds1254})
and derive a value for the period of $P = 3.933337 \pm 0.000010\,[\pm
0.000064]$ h (see Table~\ref{tbl:detect}).  This value is close to that of
\citet{diaztr09} but our estimate of the uncertainty is larger.  We do
not believe that the ASM data alone warrant a more precise estimate of
the orbital period than we have given.

\begin{figure}[tbp]
\includegraphics[height=3.5in,angle=0.0]{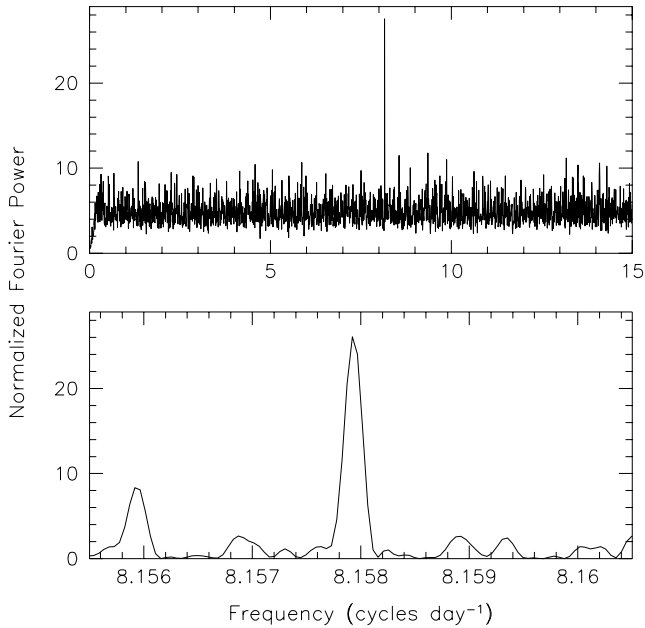}
\caption{(top) Portion of a compressed power density spectrum made
from the 1.5-12 keV light curve of 4U\,1323$-$62.  The ``wtgs'' method
and a 3.0 day smoothing time scale were employed in producing the
PDS. (bottom) A small region of the corresponding uncompressed PDS.
\label{fig:pds1323}}
\end{figure}

{\bf 4U\,1323$-$62} is also a dipping LMXB.  \citet{parmar89} report on
{\it EXOSAT} X-ray observations wherein the orbital period of the
binary system was manifest in terms of intensity dips that recurred
with a period of $P = 2.932 \pm 0.005$ h.  Later, \citet{balchch99}
derived a period of $P = 2.938 \pm 0.020$ h from X-ray observations
with {\it BeppoSAX}.  Even though the source is rather weak, the
orbital period is easily detected in the ASM light curves (see
Fig.~\ref{fig:pds1323}).  We derive a value for the period of $P =
2.941923 \pm 0.000005\,[\pm 0.000036]$ h (see Table~\ref{tbl:detect}).
It is interesting to compare this value with the conclusion of
\citet{parmar89} that consideration of the times of dips in two
observations separated by one year imply that the period must be more
precisely given by $P = 2.932128 \pm 0.000002 \pm n \times 0.000977$
h.  The ambiguity in this formula reflects an ambiguity in the number
of orbital cycles in the one year interval between the observations.
The present determination of the orbital period implies $n = 10$ and
that $P = 2.941898 \pm 0.000005$ h where the uncertainty is dominated
by the uncertainty of the correction for $n = 10$.  Given that the
difference between this period and the ASM-derived value is only
$\Delta P = 0.000025$ h and considering the uncertainties in the two
measurements, we conclude that the two values are mutually consistent.
Recent studies of this source may be found in \citet{balchch09} and
\citet{bal09}.

\begin{figure}[tbp]
\includegraphics[height=3.5in,angle=0.0]{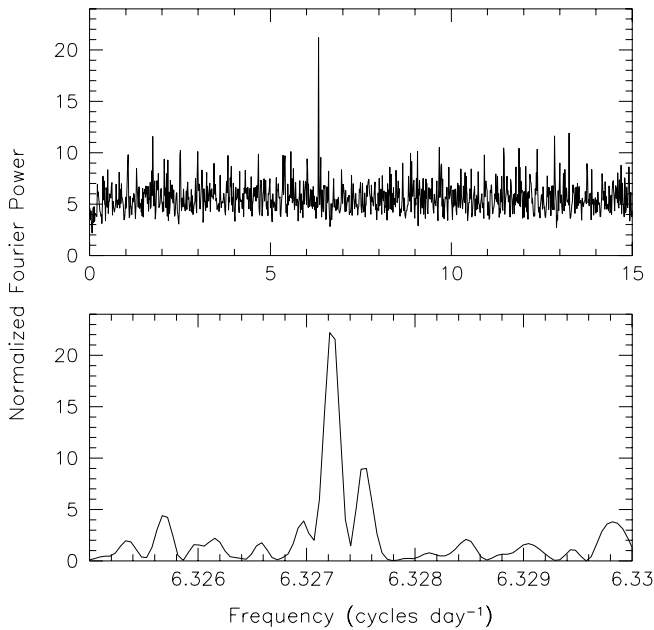}
\caption{(top) Portion of a compressed power density spectrum made
from the 1.5-12 keV light curve of 4U\,1636$-$536.  The ``wtgs'' method
and a 0.9 day smoothing time scale were employed in producing the
PDS. (bottom) A small region of the corresponding uncompressed PDS.
\label{fig:pds1636}}
\end{figure}

{\bf 4U\,1636$-$536} is a persistent LMXB that often displays Type I
X-ray bursts.  The orbital period was found through optical photometry
\citep{peder81}.  It is also evident through radial velocity
measurements made from emission line profiles obtained in optical
spectroscopic observations \citep[e.g.,][]{augus98,casar06}.
Estimates of the orbital period have been successively improved by
\citet{vp90,augus98,giles02} since the discovery by \citet{peder81}.
To our knowledge, the best optical ephemeris is that of
\citet{giles02}, who estimate that the period is $P = 3.7931263 \pm
0.0000038$ h.  The ASM detection (see Figure~\ref{fig:pds1636}) is
secure, given knowledge of the optical period, but would not be
sufficiently strong to have been detected in our blind search.  It is,
nonetheless, the first detection of the orbital period in X-rays.  The
period estimated from the ASM results is $P = 3.793128 \pm
0.000008\,[\pm 0.000060]$ h (Table~\ref{tbl:detect}).  This value is
consistent with that of \citet{giles02} though it is somewhat less
precise.

\begin{figure}[tbp]
\includegraphics[height=3.5in,angle=0.0]{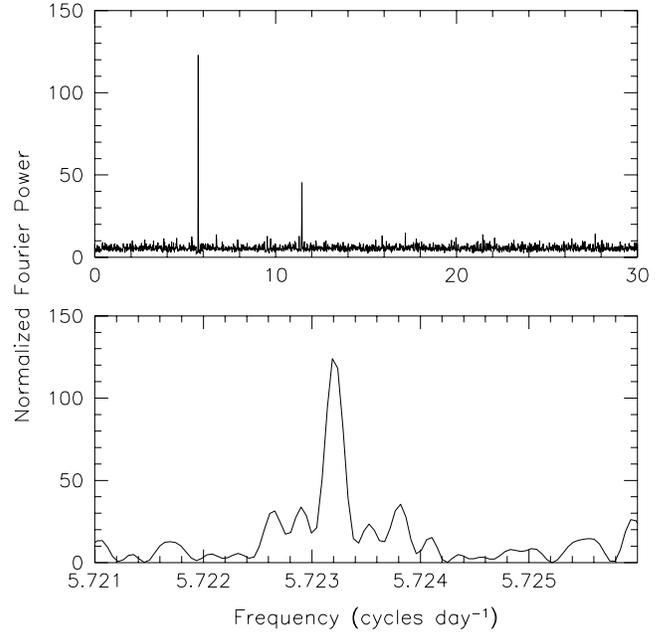}
\caption{(top) Portion of a compressed power density spectrum made
from the 1.5-12 keV light curve of GX\,9+9.  The ``wtgs'' method
and a 0.9 day smoothing time scale were employed in producing the
PDS. (bottom) A small region of the corresponding uncompressed PDS.
\label{fig:pdsgx9p9}}
\end{figure}

{\bf 4U\,1728$-$16 (GX\,9+9)} is a rather bright persistent LMXB.  The
orbital periodicity at the period of $P = 4.19$ hours was found in
{\it HEAO\,1} A-1 X-ray data by \citet{hw88}.  Soon afterwards,
\citet{s90} found variations at essentially the same period in the
brightness of the optical counterpart.  The ASM detection and an
account of the variability in the ASM data of the modulation amplitude
is described in full by \citet{harris09}.  The periodicity is manifest
with even greater significance in the longer data set used in our most
recent analyses; compare Fig.~\ref{fig:pdsgx9p9} with Fig.~2 of
\citet{harris09}.  In Table~\ref{tbl:detect} we also give an updated
estimate of the orbital period that is only slightly different from
that given by \citet{harris09}.

\begin{figure}[tbp]
\includegraphics[height=3.5in,angle=0.0]{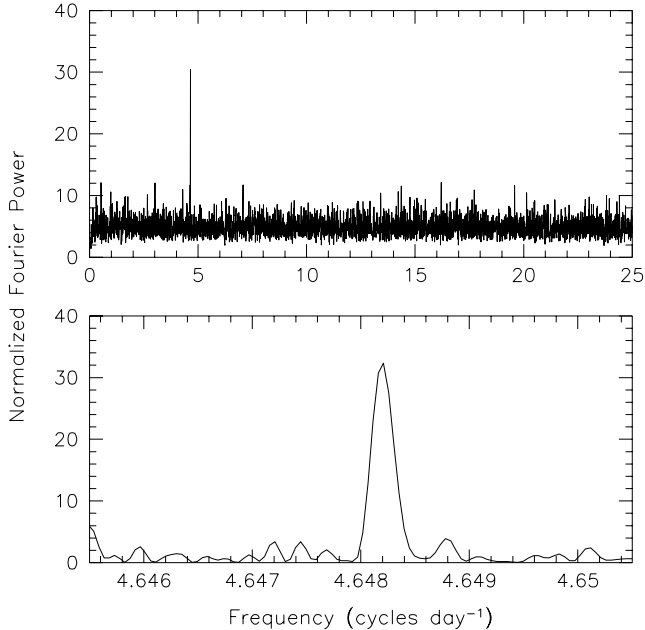}
\caption{(top) Portion of a compressed power density spectrum made
from the 1.5-12 keV light curve of 4U\,1746$-$37.  The ``wtgs-v8'' method
and a 3.0 day smoothing time scale were employed in producing the
PDS. (bottom) A small region of the corresponding uncompressed PDS.
\label{fig:pds1746}}
\end{figure}

{\bf 4U\,1746$-$37} is an LMXB in the globular cluster NGC 6441. X-ray
observations carried out with {\it EXOSAT}, {\it Ginga}, and {\it
RXTE}/PCA showed the presence of dips that occur roughly every 5 hours
and quickly led to the interpretation that their recurrence period is
the orbital period \citep{parm89,sansom93,bc2004}.  The current best
estimate of the orbital period, $P = 5.16 \pm 0.01$ h, was given by
\citet{bc2004}.  This periodicity is seen with high significance in
the ASM data (see Fig.~\ref{fig:pds1746}).  We obtain the period
estimate $P = 5.163287 \pm 0.000018\,[\pm 0.000111]$ h that is
consistent with the results of \citet{bc2004} but is substantially
more precise.

\begin{figure}[tbp]
\includegraphics[height=3.5in,angle=0.0]{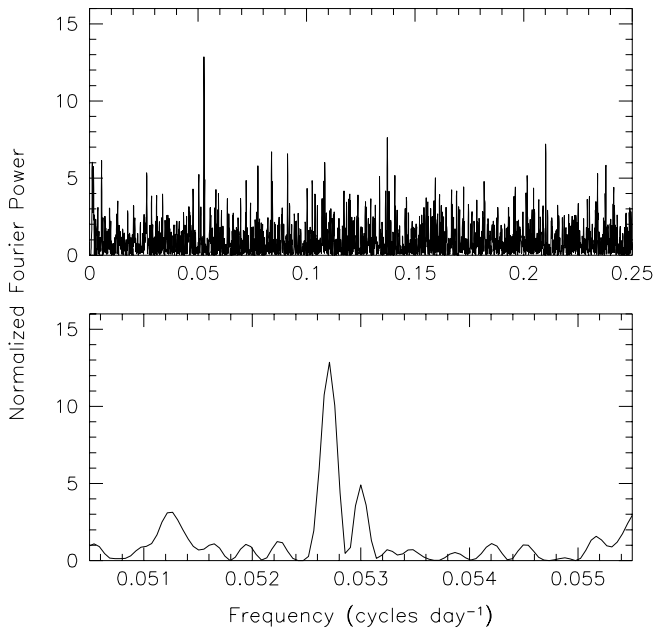}
\caption{(top) Power density spectrum made
from the 3-5 keV light curve of GRS\,1758$-$258.  The ``wtgs'' method
and a 500.0 day smoothing time scale were employed in producing the
PDS. (bottom) A small region of the uncompressed PDS.
\label{fig:pds1758}}
\end{figure}

{\bf GRS\,1758$-$258} most likely comprises a black hole accreting
from a K-type giant star \citep[and references
therein]{smithetal02,rothetal02}.  \citet{smithetal02} found a
periodicity with $P = 18.45 \pm 0.01$ days in the data obtained in
extensive {\it RXTE} PCA observations done in the years 1997 through
2000 and interpreted the period as the orbital period.
\citet{rothetal02} performed near-IR observations and identified an
early K-type star as the likely companion if the orbital period is as
long as 18.45 days.  We find a peak in the ASM power spectrum shown in
Figure~\ref{fig:pds1758} that corresponds to a period of $P = 18.973
\pm 0.007\,[\pm 0.036]$ days and that we therefore identify with the
orbital period.

\begin{figure}[tbp]
\includegraphics[height=3.5in,angle=0.0]{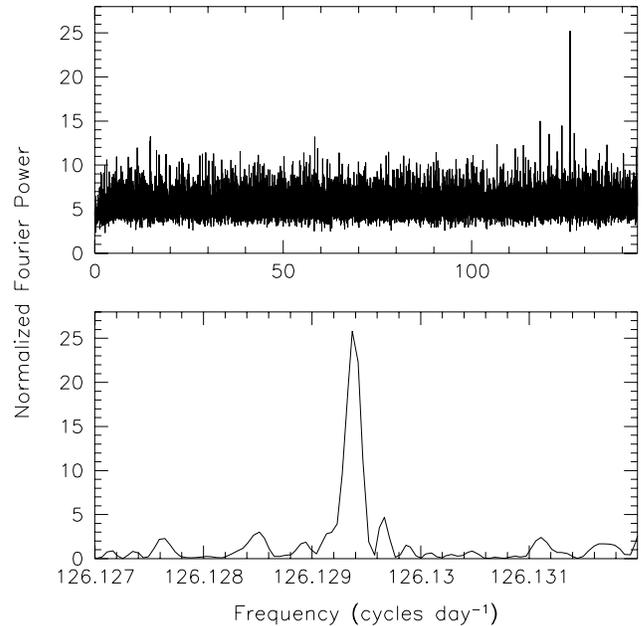}
\caption{(top) Portion of a compressed power density spectrum made
from the 1.5-12 keV light curve of 4U\,1820$-$30.  The ``wtgs'' method
and a 0.9 day smoothing time scale were employed in producing the
PDS. (bottom) A small region of the corresponding uncompressed PDS.
\label{fig:pds1820}}
\end{figure}

{\bf 4U\,1820$-$30} is a very bright LMXB in the globular cluster NGC
6624.  \citet{spw87} analyzed data from X-ray observations with {\it
  EXOSAT} and discovered the 11 minute orbital period.  The
periodicity is only evident as a 3\% or less peak-to-peak modulation
in the overall X-ray intensity.  \citet{chougrn01} used measurements
made by a number of X-ray satellites of the times of maxima of the
11-m variation to obtain an ephemeris wherein both linear and
quadratic terms are significant.  To our knowledge, this is the most
precise ephemeris in the literature at this time.  We detect the
periodicity in the ASM light curve with high significance; see
Fig.~\ref{fig:pds1820} and Table~\ref{tbl:detect}.  The period we
obtain, $P = 685.01092 \pm 0.00007\,[\pm 0.00054]$ s, is an average
over the duration of the ASM data set and effectively applies near the
mid-time of this interval, i.e., near MJD 52620.  We can compare our
period measurement with the period of $P = 685.01126$ s predicted for
this epoch by the \citet{chougrn01} ephemeris.  If we assume that the
smaller of the two uncertainties in the ASM period that are given in
Table~\ref{tbl:detect} is applicable, then the ASM period is about 4
standard deviations below the prediction of the \citet{chougrn01}
ephemeris.  This suggests that the coefficient of the quadratic term
in that ephemeris should be more negative.  However, we have not
undertaken a proper joint analysis of all of the timing results and so
this conclusion must be regarded as tentative.

\subsection{Additional Detections of High-Mass X-ray Binaries}

The orbital periods of a number of HMXBs are listed in
Table~\ref{tbl:detect}.  In contrast with the results on LMXBs,
Table~\ref{tbl:detect} shows that the majority of the most significant
detections of these systems were made in the 5-12 keV energy band.
This is in accordance with expectations based on the observation that
many of the accreting compact objects in HMXBs are X-ray pulsars that
tend to have relatively hard X-ray spectra.

{\bf A\,0535+26} is a Be/X-ray pulsar that can be rather bright in
X-rays during transient outbursts. The early observations are reviewed
by \citet{fwh96}.  Pulse timing analyses of observations obtained with
the BATSE instrument were used to determine that the orbital period is
$P = 110.3 \pm 0.3$ days \citep[and references therein]{fwh96}.  A
portion of an ASM power density spectrum is shown in
Figure~\ref{fig:pdsj1008}.  This figure demonstrates that the
periodicity is apparent, but is not detected with sufficient
significance to allow us to obtain a useful improvement in the
precision of the orbital period.  Seven separate peaks in the
intensity of this source are apparent in the latter part of the ASM
light curve.  Their average separation is close to 114 days and is
probably not consistent with the period determined through pulse
timing.  The outbursts of some other Be star/neutron star binaries do
not occur at precisely the orbital periods determined by pulse timing,
so there is no reason in this case to doubt the accuracy of the pulse
timing value.

\begin{figure}[tbp]
\includegraphics[height=3.5in,angle=0.0]{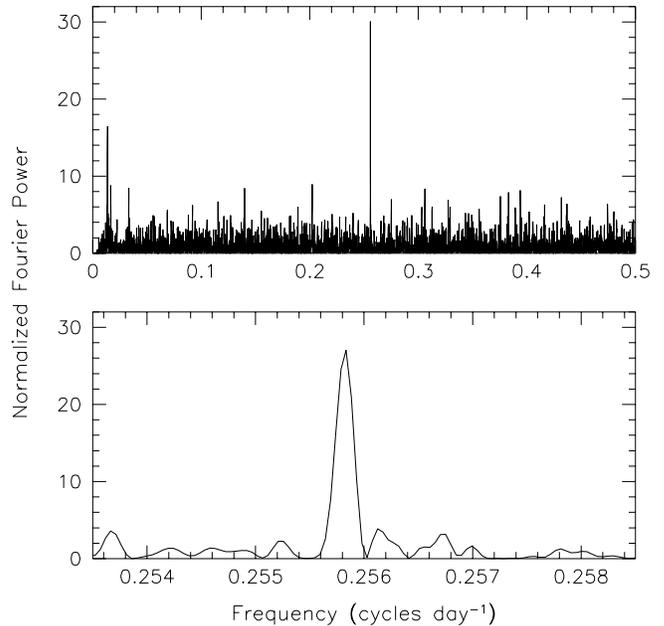}
\caption{(top) Portion of a compressed power density spectrum made
from the 1.5-12 keV light curve of LMC X-1.  The ``wtgs'' method
and a 100.0 day smoothing time scale were employed in producing the
PDS. (bottom) A small region of the corresponding uncompressed PDS.
\label{fig:pdslmcx1}}
\end{figure}

\begin{figure}[tbp]
\includegraphics[height=3.5in,angle=0.0]{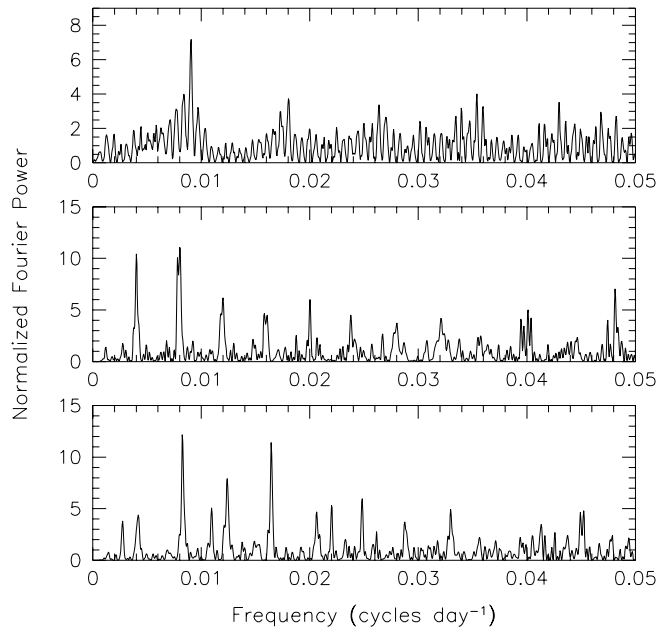}
\caption{Power density spectra made using the ``wtgs'' method and a
  500.0 day smoothing time scale. Only the low frequency regions of
  the power spectra are shown. (top) Power density spectrum made from
  the 1.5-12 keV light curve of A\,0535$+$262.  The ``ringing''
  appearance derives from a light curve where a small number of strong
  outbursts punctuate the otherwise low source strength. (middle)
  Power density spectrum made from the 3-5 keV light curve of
  GRO\,J1008$-$57. (bottom) Power density spectrum made from the 5-12
  keV light curve of 2S\,1845$-$024. The previously determined orbital
  frequencies are 0.009066\,(25) d$^{-1}$ (A\,0535$+$262),
  0.004018\,(8) d$^{-1}$ (GRO\,J1008$-$57), and 0.0041292\,(2)
  d$^{-1}$ (2S\,1845$-$024).
\label{fig:pdsj1008}}
\end{figure}

{\bf LMC X-1} is a luminous HMXB in the Large Magellanic Cloud which
highly likely comprises a stellar-mass black hole and its normal
OB-type companion. The properties of the system are described in
detail by \citet{orosz09}.  \citet{orosz09} also present the ASM
results and how they relate to other observations including other
period determinations.  \citet{orosz09} adopt the value of $3.90917
\pm 0.00005$ days as their best estimate of the period based on
optical photometry and spectroscopy. An updated ASM power spectrum is
shown in Figure~\ref{fig:pdslmcx1} and revised estimates, based solely
on the ASM power spectrum, of the orbital frequency and period are
given in Table~\ref{tbl:detect}. We note that the ASM detection
provides the only reported evidence to date of the orbital period in
X-rays and that our X-ray-based period estimate of $P = 3.90898 \pm
0.00021 [\pm 0.00153]$ days is fully consistent with the
optically-based value of \citet{orosz09}.

{\bf GRO\,J1008$-$57} is a transient X-ray pulsar in a Be/X-ray binary
discovered with the BATSE experiment on the {\it Compton Gamma-Ray
Observatory} \citep[see references in][]{shrad99,coe07}.
\citet{shrad99} used $\sim2$ years of ASM data to obtain an early
estimate of the outburst period of $P \sim 135$ d.  \citet{lcatel06}
reported a relatively precise determination of the outburst period,
$P_{outburst} = 248.9 \pm 0.5$ d, that was based on the time intervals
between widely spaced outbursts.  We believe this is still the best
current estimate of the outburst cycle time.  The ASM power spectrum
is shown in the top panel of Figure~\ref{fig:pdsj1008}.  Though the
peaks due to the outburst periodicity are clear, they are not
sufficiently well-defined to yield a superior period estimate.
\citet{coe07} have estimated the orbital period through timing of the
93-s pulsations seen in the BATSE data.  They obtained the result $P =
247.8 \pm 0.4$ d and noted that is it is in good agreement with the
result of \citet{lcatel06}.

\begin{figure}[tbp]
\includegraphics[height=3.5in,angle=0.0]{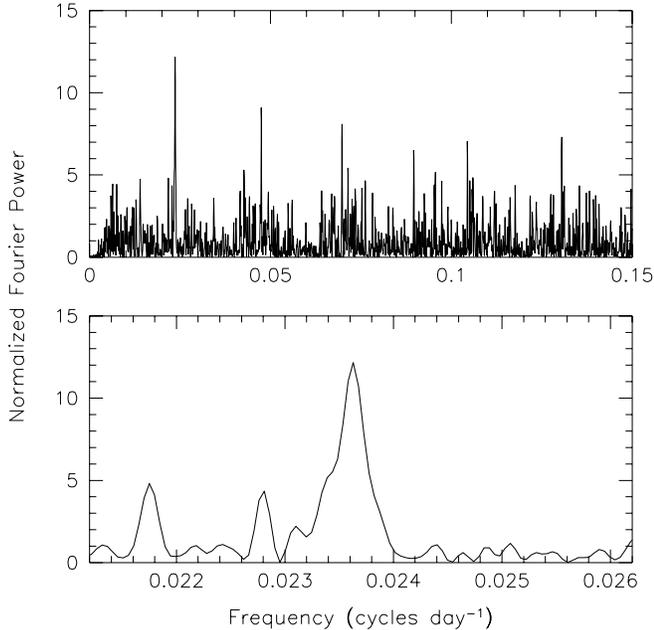}
\caption{(top) Portion of a compressed power density spectrum made
from the 5-12 keV light curve of 2S\,1417$-$62.  The ``wtgs'' method
and a 100.0 day smoothing time scale were employed in producing the
PDS. (bottom) A small region of the corresponding uncompressed PDS.
The orbital frequency determined through pulse timing is 0.02374 (2)
d$^{-1}$ (see text).
\label{fig:pds1417}}
\end{figure}

{\bf 2S\,1417$-$62} is also a transient pulsar in a Be/X-ray binary
system. A brief description of the early history and of a pulse-timing
analysis using BATSE data may be found in \citet{fwc96}.  The BATSE
timing analysis yielded an estimate of the orbital period of $P =
42.12 \pm 0.03$ d as well as the projected semimajor axis,
eccentricity, epoch of periastron passage, and other orbital elements.
A slight revision to the orbital period and time of periastron passage
have been given by \citet{inam04}.  Evidence of outbursts recurring at
intervals that are close in duration to the orbital period (or
multiples thereof) is seen in the ASM power spectrum (see
Fig.~\ref{fig:pds1417}).  Frequency and period estimates obtained from
the spectrum shown in Fig.~\ref{fig:pds1417} are listed in
Table~\ref{tbl:detect} but are not as precise as the pulse-timing
values.

\begin{figure}[tbp]
\includegraphics[height=3.5in,angle=0.0]{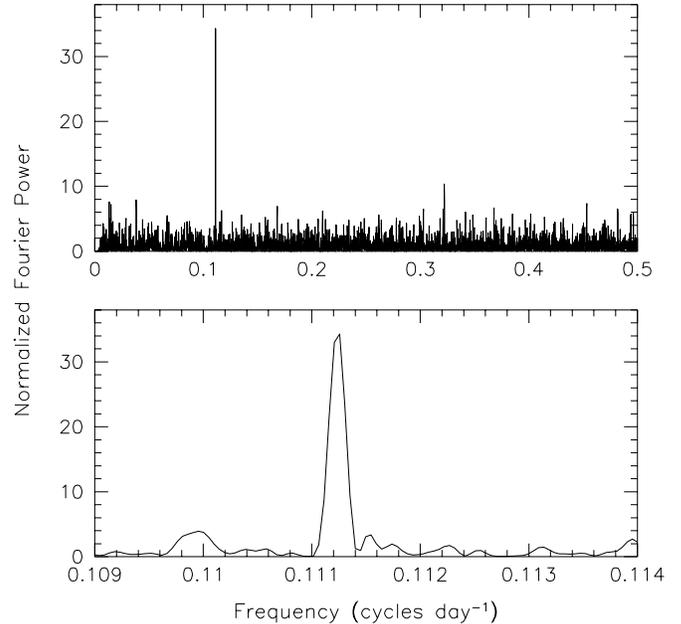}
\caption{(top) Portion of a compressed power density spectrum made
from the 5-12 keV light curve of IGR\,J16320$-$4751.  The ``wtgs'' method
and a 100.0 day smoothing time scale were employed in producing the
PDS. (bottom) A small region of the corresponding uncompressed PDS.
\label{fig:pds1632}}
\end{figure}

{\bf IGR\,J16320$-$4751} is an X-ray pulsar with the rather long pulse
period of $\sim 1300$ s \citep{lutov05}.  \citet{corbetetal05} found
the $8.96 \pm 0.01$ day orbital period by analyzing a {\it Swift}
Burst Alert Telescope 14-200 keV light curve, and determined the epoch
of maximum flux to be MJD $53507.1 \pm 0.1$. \citet{walteretal06} used
observations with {\it INTEGRAL} to measure the period, obtaining $P =
8.99 \pm 0.05$ d.  The orbital period is seen at a high level of
significance in the ASM power spectrum shown in
Figure~\ref{fig:pds1632}.  We use the ASM power spectrum to estimate
that the period is $8.9901 \pm 0.0009\,[\pm 0.0081]$ d.

\begin{figure}[tbp]
\includegraphics[height=3.5in,angle=0.0]{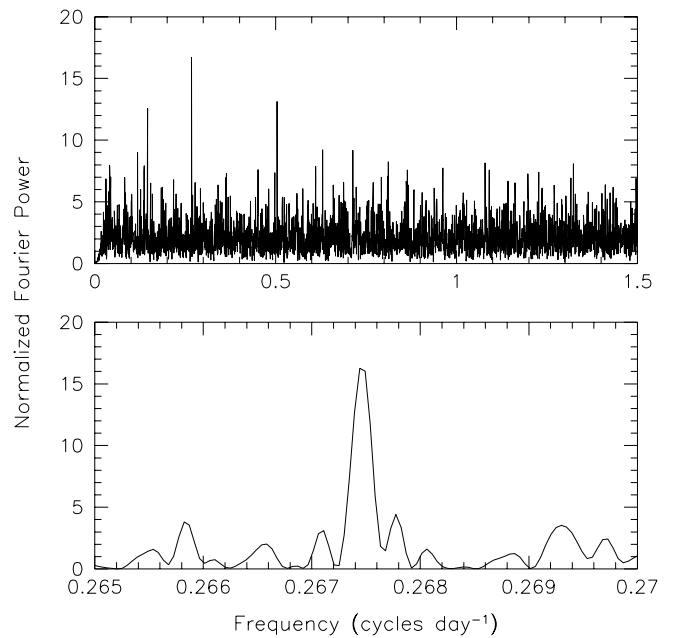}
\caption{(top) Portion of a compressed power density spectrum made
from the 5-12 keV light curve of IGR\,J16418$-$4532.  The ``wtgs''
method and a 30.0 day smoothing time scale were employed in producing
the PDS. (bottom) A small region of the corresponding uncompressed
PDS.
\label{fig:pds1641}}
\end{figure}

{\bf IGR\,J16418$-$4532} is an X-ray source in the Galactic plane that
was discovered using {\it INTEGRAL} \citep{tomsicketal04}.
\citet{corbetal06} found and measured the orbital period using both
{\it Swift}/BAT and ASM data and obtained the values $3.753 \pm 0.004$
days and $3.7389 \pm 0.0004$ days, respectively.  Our current ASM
power spectrum (Fig.~\ref{fig:pds1641}) shows a peak that is
moderately significant given prior knowledge of the period.  The
frequency of this peak corresponds to the period $P = 3.73886 \pm
0.00028\,[\pm 0.00140]$ days.  This period is essentially identical to
that determined earlier from the ASM data by \citet{corbetal06}.

\begin{figure}[tbp]
\includegraphics[height=3.5in,angle=0.0]{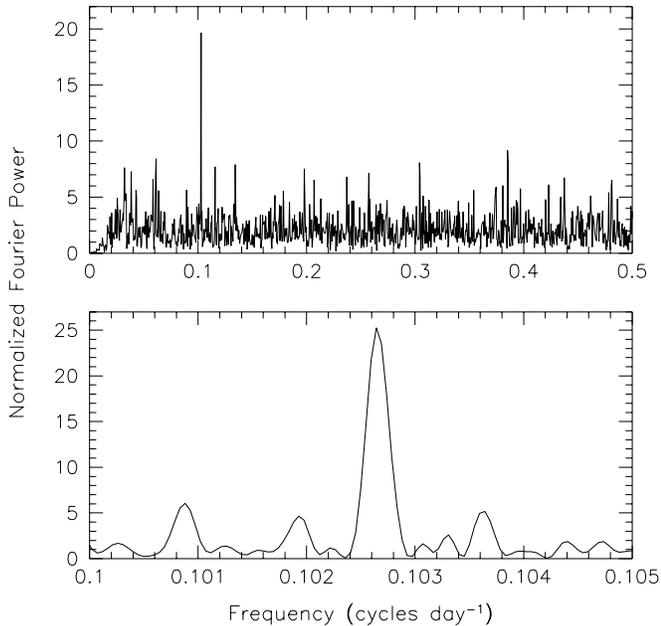}
\caption{(top) Portion of a compressed power density spectrum made
from the 5-12 keV light curve of EXO\,1722$-$363.  The ``wtgs'' method
and a 30.0 day smoothing time scale were employed in producing the
PDS. (bottom) A small region of the corresponding uncompressed PDS.
\label{fig:pds1722}}
\end{figure}

{\bf EXO\,1722$-$363} is an X-ray pulsar in a high-mass binary system
seen edge-on \citep[and references therein]{thompson07}.  Eclipses
were found in X-ray observations made in scans of the Galactic plane
with the {\it RXTE} PCA and their temporal spacing was used to get the
first real measurement of the orbital period, i.e., $P = 9.741 \pm
0.004$ days \citep{cms05}.  \citet{thompson07} have estimated the
orbital parameters through a pulse timing analysis and obtain a value
for the period of $P = 9.7403 \pm 0.0004$ days.  One might expect a
strong signal from an eclipsing system but the source intensity is
rather modest, about 4 mcrab (1.5 - 12 keV) on average in the ASM
observations.  Thus only a modest peak is evident close to the
frequency in the ASM power spectrum (see Figure~\ref{fig:pds1722}).
We estimate the period to be $P = 9.7414 \pm 0.0018\,[\pm 0.0095]$
days.  This is consistent with but somewhat less precise than the
pulse-timing-based estimate of \citet{thompson07}.
\citet{masonetal09} have recently reported estimates of the masses of
the two components of the binary system and confirm that it is indeed
an HMXB.

\begin{figure}[tbp]
\includegraphics[height=3.5in,angle=0.0]{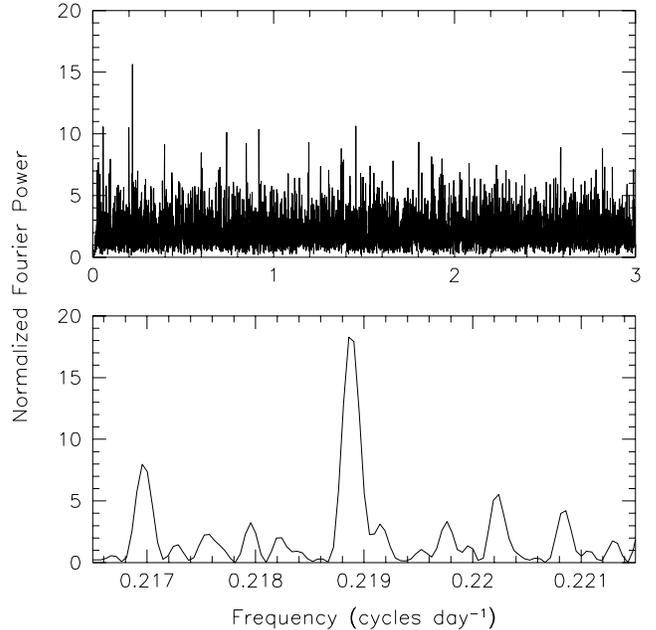}
\caption{(top) Portion of a compressed power density spectrum made
from the 5-12 keV light curve of IGR\,J18027$-$2016.  The ``wtgs'' method
and a 30.0 day smoothing time scale were employed in producing the
PDS. (bottom) A small region of the corresponding uncompressed PDS.
\label{fig:pds18027}}
\end{figure}

{\bf SAX\,J1802.7$-$2017} = {\bf IGR\,J18027$-$2016} is only
$\sim22\arcmin$ away from the bright X-ray source GX\,9+1.  It was
discovered using {\it BeppoSAX} and, despite its proximity to a much
brighter source, was found to be an X-ray pulsar in a high-mass X-ray
binary system \citep{aug03}.  Changes in the apparent pulse period led
\citet{aug03} to infer that the pulsar is in an orbit with a period of
$\sim4.6$ days.  \citet{hilletal05} realized that the {\it INTEGRAL}
source IGR\,J18027$-$2016 is associated with the SAX source and
refined the orbital period to be $P = 4.5696 \pm 0.0009$
days. \citet{jainetal09} used {\it Swift}-BAT, {\it INTEGRAL}-ISGRI,
and {\it RXTE}-ASM data to estimate that the orbital period is $P =
4.5693 \pm 0.0004$ days.  This periodicity is apparent in our ASM
power spectra (see Fig.~\ref{fig:pds18027}).  We use the present
results to estimate $P = 4.5687 \pm 0.0003\,[\pm 0.0021]$ days.

{\bf 2S\,1845$-$024} = {\bf GS\,1843$-$02} is an accreting pulsar that
is evident during widely spaced outbursts \citep[and references
therein]{fingeretal99}.  A fairly precise value for the orbital period
of $P = 242.18 \pm 0.01$ days was determined by \citet{fingeretal99}
by a pulse-timing analysis of BATSE data.  A small number, $\sim 5$,
of outbursts are evident in the mission-long ASM light curve and the
outburst periodicity is weakly evident in the ASM power spectrum (see
the lower panel of Figure~\ref{fig:pdsj1008}).  However, the ASM
results do not support a determination of an orbital period that can
approach the precision of the pulse-timing-based value determined by
\citet{fingeretal99}.  Hence, we do not present an ASM-based estimate
of the period.

\begin{figure}[tbp]
\includegraphics[height=3.5in,angle=0.0]{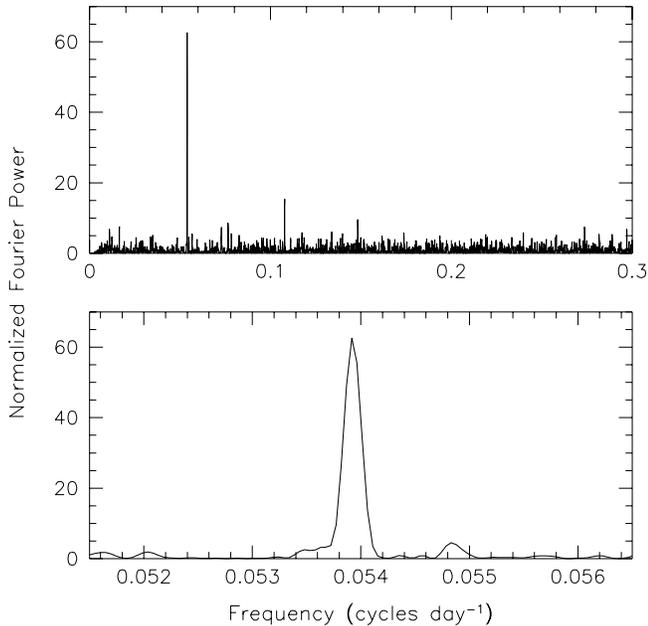}
\caption{(top) Portion of a compressed power density spectrum made
from the 5-12 keV light curve of IGR\,J18483$-$0311.  The ``wtgs'' method
and a 100.0 day smoothing time scale were employed in producing the
PDS. (bottom) A small region of the corresponding uncompressed PDS.
\label{fig:pds1848}}
\end{figure}

{\bf IGR\,J18483$-$0311} was discovered in {\it INTEGRAL}/IBIS
observations of the Galactic Center region \citep{chern03}.  The
orbital period was independently found around the same time by
\citet{lcatel06} in the ASM light curve and by \citet{sguera07} in the
{\it INTEGRAL} light curve. The reported values for the period were
$18.55 \pm 0.03$ days and $18.52 \pm 0.01$ days, respectively.
\citet{jainetal09} have recently analyzed {\it Swift}/BAT and ASM data
and estimate the period to be $P = 18.5482 \pm 0.0088$ days.  The
period is detected with high significance in the present study (see
Figure~\ref{fig:pds1848}).  We use that power spectrum to estimate
that the orbital period is $P = 18.545 \pm 0.003\,[\pm 0.034]$ days.

\begin{figure}[tbp]
\includegraphics[height=3.5in,angle=0.0]{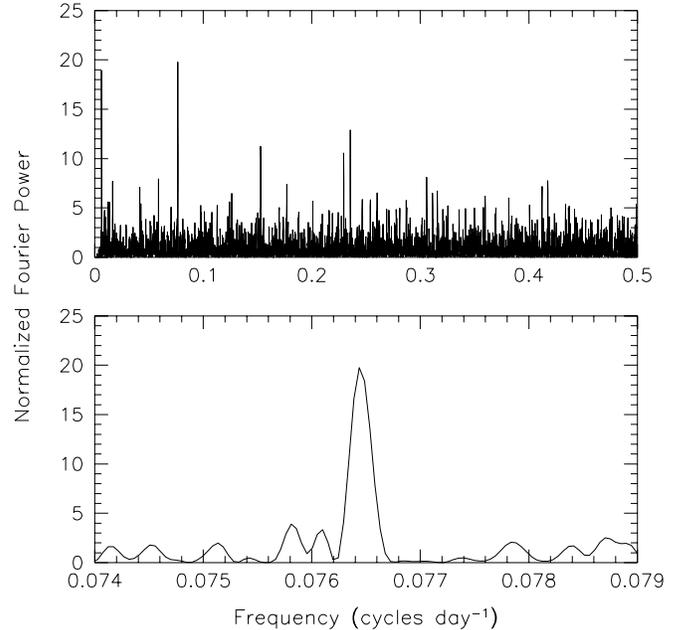}
\caption{(top) Portion of a compressed power density spectrum made
from the 5-12 keV light curve of SS\,433.  The ``wtgs'' method
and a 100.0 day smoothing time scale were employed in producing the
PDS. (bottom) A small region of the corresponding uncompressed PDS.
\label{fig:pds433}}
\end{figure}

{\bf 3A\,1909+048 } is the X-ray counterpart to the well-known object
{\bf SS\,433}.  This object most likely consists of a stellar-mass
black hole that is accreting at super-Eddington rates from a strong
wind emitted by a high-mass normal-type star, and produces jets
wherein the outflow velocity is approximately $0.26 c$.  An early
review may be found in \citet{margon84} and a more recent one in
\citet{fabrika04}.  The precession of the jets with a period of $\sim
162$ days is not only evident in the moving optical lines, but also in
other phenomena including the apparent X-ray intensity
\citep[see][]{wen06,giesetal02}.  A number of reports discuss the
evidence from optical spectroscopy for Doppler shifts of the compact
and normal components and their interpretation in terms of orbital
parameters and component masses \citep[see,
e.g.,][]{hillwigetal08,cheretal09,kubetal10}.  \citet{goranetal98}
reported a precise measurement of the orbital period of the binary,
i.e., $P_{orb} = 13.08211 \pm 0.00008$ days, on the basis of the times
of eclipse-like dips in the optical flux. The X-ray intensity also
shows eclipse-like dips \citep[e.g.,][]{filipetal06,cheretal09} but
these have not yet been exploited for orbital period estimation.
\citet{giesetal02} reported the first evidence of the orbital period
in the ASM X-ray intensity measurements.  Very recently,
\citet{kubetal10} obtained a new time of minimum optical brightness
and used it, together with the earlier observations, to obtain a
revised value of the period, $P_{orb} = 13.08227 \pm 0.00008$ d.  In
the present analysis, we find unambiguous evidence of both the
precession and the orbital periods in the power spectrum (see
Figure~\ref{fig:pds433}).  Our estimate of the orbital period is
$P_{orb} = 13.080 \pm 0.003\,[\pm 0.017]$ days.  This value is
consistent with, but somewhat less precise than, those of
\citet{goranetal98} and \citet{kubetal10}. If we fold the ASM light
curve with the period and epoch of \citet{kubetal10}, we find that the
time of X-ray minimum is coincident with the time of optical
minimum. The shape of the folded ASM light curve is similar to the
eclipse-like shapes seen by {\it INTEGRAL} \citep[e.g.,][]{cheretal09}
and suggests that the eclipse-like decrease in the flux is the result
of periodically occurring partial occultations of the X-ray emitting
regions, broadly construed to include regions where the X-ray flux may
be scattered (see Fig.~\ref{fig:fold2}).

\begin{figure}[tbp]
\includegraphics[height=3.5in,angle=0.0]{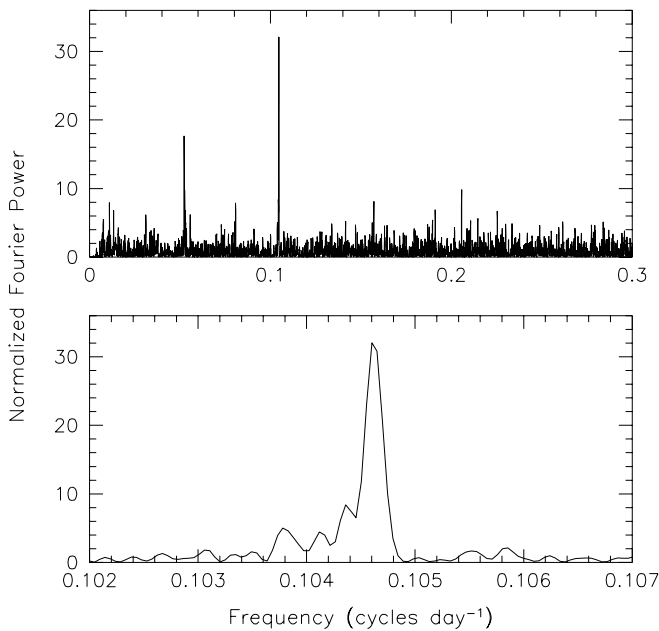}
\caption{(top) Portion of a compressed power density spectrum made
from the 1.5-12 keV light curve of 4U\,2206+543.  The ``wtbx'' method
and a 100.0 day smoothing time scale were employed in producing the
PDS. (bottom) A small region of the corresponding uncompressed PDS.
\label{fig:pds2206}}
\end{figure}

{\bf 4U\,2206+543} is a high-mass X-ray binary which manifests a
distinct variation at a period of $P \approx 9.6$ days in the ASM
light curve \citep[][]{corbetpeele01}.  This source is included in
this report even though the $\sim 9.6$-day period was reported by
\citet{wen06} because the underlying period may actually be twice as
long, or $\sim 19.2$ days (\citealt{corbmt07}; see also
\citealt{wang09}).  In this regard, \citet{corbmt07} found a signal at
a period of $P = 19.25 \pm 0.08$ days in {\it Swift}/BAT data and a
signal near the same period in ASM data.  \citet{wang09} found a peak
near a period $P= 19.11$ days in the ASM data.  Our power spectrum of
ASM data also shows peaks at frequencies corresponding to both periods
(see Figure~\ref{fig:pds2206}).  The more significant peak corresponds
to a period of $P = 9.5584 \pm 0.0012\,[\pm 0.0091]$ days.  Assuming
that the subharmonic corresponds to a period twice this value, we
obtain $P = 19.117 \pm 0.003\,[\pm 0.018]$ days.  Radial velocity
measurements from optical or near IR spectroscopy are needed to
unambiguously determine the true orbital period.

\subsection{Nondetections and Search Sensitivity}

\begin{deluxetable*}{lcccccc}
\tablecolumns{7}
\tablewidth{0pt}
\tabletypesize{\footnotesize}
\tablecaption{Upper Limits on Selected Periodicities\label{tbl:uplim}}
\tablehead{
\colhead{Source Name} &
\colhead{Band} &
\colhead{Intensity\tablenotemark{a}} &
\colhead{Period\tablenotemark{b}} &
\colhead{Filter Time\tablenotemark{c}} &
\colhead{Amplitude\tablenotemark{d}} &
\colhead{A/I\tablenotemark{e}}
 \\
\colhead{} &
\colhead{(keV)} &
\colhead{(cts s$^{-1}$)} &
\colhead{} &
\colhead{(days)} &
\colhead{(cts s$^{-1}$)} &
\colhead{}
 }
\startdata

Crab Nebula & 1.5-12 & 75.50\,(1) & & 3.0 & 0.12 & 0.0016 \\
Sco\,X-1 & 1.5-12 & 867.81\,(6) & 0.789 d & 3.0 & 7.2 & 0.0082 \\
Sco\,X-1 & 1.5-12 & 867.81\,(6) & 0.789 d & 10.0 & 8.5 & 0.0099 \\
GX\,340$+$0 & 1.5-12 & 28.657\,(8) & & 3.0 & 0.36 & 0.013 \\
GRO\,J1655$-$40 & 1.5-12 & 5.311\,(7) & 2.62 d & 30.0 & 0.30 & 0.056 \\
GX\,339$-$4 & 1.5-12 & 4.337\,(6) & 1.75 d & 10.0 & 0.083 & 0.019 \\
4U\,1735$-$444 & 1.5-12 & 12.739\,(7) & 4.65 h & 3.0 & 0.11 & 0.0089 \\
GX\,5$-$1 & 1.5-12 & 69.30\,(1) & & 3.0 & 0.75 & 0.011 \\
V4641 Sgr & 1.5-12 & 0.458\,(9) & 2.80 d & 30.0 & 0.13 & 0.28 \\
4U\,1957+115 & 1.5-12 & 2.366\,(4) & 9.33 h & 3.0 & 0.040 & 0.017 \\

\enddata

\tablenotetext{a}{Weighted average of the entire ASM light curve.  See
note (a) to Table~\ref{tbl:ampl}.}

\tablenotetext{b}{Approximate orbital period if reported.  See the
following references for previous period determinations and more
precise values: \citealt{vlr03} (Sco X-1); \citealt{gbo01}
(GRO\,J1655$-$40); \citealt{hynesetal03} (GX339$-$4);
\citealt{casar06} (4U\,1735$-$444); \citealt{oroetal01} (V4641 Sgr);
\citealt{thoretal87} (4U\,1957+115);}

\tablenotetext{c}{See text and column 1 of Table~\ref{tbl:tmscls}.}

\tablenotetext{d}{Amplitude which would have been likely to yield a
peak of $\sim25$ in the power density spectrum.}

\tablenotetext{e}{Ratio of the signal amplitude that would yield a
peak power of 25 (A; from column 6) to the average source intensity
(I; from column 3).}

\end{deluxetable*}

The estimates of the amplitudes of the periodic signals that are
reported in Table~\ref{tbl:ampl} fall, for most of the cases, just
below the detection thresholds of the present blind search.  They
thereby illustrate the typical sensitivity of the present search.  We
have also estimated the amplitudes that approximately correspond to
our detection thresholds in a set of additional sources chosen either
to help cover a wide range of intensities or because they are of
interest for other reasons.  To accomplish this we have
computationally superimposed sine waves onto the ASM light curves and
repeated the search analyses on the light curves for those particular
sources in order to determine the signal amplitudes that would have
resulted in peaks reaching renormalized whitened powers of $\sim 25$
in the power density spectra.  These estimates, which may be taken as
upper limits since we would have recognized peaks near these
frequencies that reached powers as small as 15, are listed in
Table~\ref{tbl:uplim}.  This table also lists previously reported
periods for some of the sources as well as mission-average
intensities.  Column 7 of the table gives the effective upper limits
on the amplitudes as fractions of the average source intensities.

\citet{lcatel06} reported modest significance detections of
periodicities in ASM power density spectra for GRO J1655$-$40,
GX\,339$-$4, V4641 Sgr, and 4U\,1957+115 at or very close to the
periods previously reported in the literature.  In the analyses using
the entire ASM light curves that we have done recently we do not
confirm these possible detections.  Results of sensitivity analyses
for these sources are included in Table~\ref{tbl:uplim}.

Table~\ref{tbl:uplim} also includes entries for two bright sources,
Sco X-1 and 4U\,1735$-$444, with well-established periodicities that
we do not detect as well as entries for three sources, i.e., the Crab
Nebula, GX\,340+0, and GX\,5$-$1, that do not have known orbital
periodicities.  Table~\ref{tbl:uplim} gives upper limits, calculated
as described above, on any modulation near the reported periods.  We
note that for Sco X-1 the noise is dominated by the intrinsic source
variations.  In rare cases, such as that described by \citet{vlr03},
specialized techniques may be available to partially mitigate the
effects of the intrinsic variation and improve the sensitivity of a
periodicity search.  We have not tried to apply any such techniques in
the present general search.  The upper limits on periodicities in the
power spectra for the Crab, GX\,340+0, and GX\,5$-$1 are included in
this table to illustrate the typical sensitivity of the present search
in the cases of bright sources.

We also did not detect the alias of the 290-s pulse period in
X1145$-$619 that was reported in \citet{lcatel06}.  This is not so
unexpected since pulse periods tend to change more rapidly than
orbital periods and could change to a degree that makes them
significantly noncoherent and therefore hard to detect in a 14-yr data
set.

In short, Tables~\ref{tbl:ampl} and \ref{tbl:uplim} show that the
search sensitivity depends on a number of factors including source
strength, degree of intrinsic source variability, and quality of the
ASM light curve (which, in turn depends on the source location
relative to nearby bright sources, etc.).

\section{Discussion}

The present search for periodicities in the ASM light curves has
proved to be more sensitive than the earlier search done by
\citet{wen06}.  The weighting and filtering aspects of our search
algorithms have proved to be major factors in the sensitivity
improvements. Another major factor is the use of light curves that
cover $\sim$5 more years.  The result is the detection of the orbital
periods in 8 LMXBs and in $\sim$12 HMXBs that were not detected by
\citet{wen06}.  Our detection of the 18.55-d period of
IGR\,J18483$-$0311 represents the co-discovery with \citet{sguera07}
of the orbital period of this system.  Our detections of the orbital
periods of 4U\,1636$-$536 and LMC X-1 are the first for these systems,
to our knowledge, that have been accomplished on the basis of X-ray
observations.

It is interesting to compare the present list of detections together
with those of \citet{wen06} with Tables 1.1 and 1.3 of
\citet{whiteetal95} that list the well-established orbital periods of
LMXBs and HMXBs as of 1994 or 1995.  Table 1.1 of \citet{whiteetal95}
lists 32 LMXBs with known orbital periods.  Of these, seven have been
in quiescence through the entire RXTE mission and one (CAL 87) is a
supersoft source that emits at energies below the ASM band; these
systems would not be expected to yield periodicity detections in ASM
data. The periodicities of 17 of the remaining 24 systems have now
been detected in the ASM light curves if we include Sco~X-1 per the
results of \citet{vlr03} even though its orbital period was not
detected in the present study.  Seven systems that are detected by the
ASM by virtue of their intensities do not yield period detections in
the present search. The ASM periodicity searches have yielded clear
detections of the orbital periods of 13 of the 16 systems listed in
Table 1.1 of \citet{whiteetal95} that had, at that time, shown
evidence of the orbital period in X-rays (see column 5 therein).  Two
of the remaining three systems have been in quiescence throughout the
RXTE mission.  The last remaining system is Cyg X-2 for which we find
no evidence of the 9.8-d period even though it is rather bright in
X-rays.

Table 1.3 of \citet{whiteetal95} lists 24 HMXBs with known orbital
periods.  Of these, only one (X0535$-$668) has been in quiescence
during the entire RXTE mission. The orbital periodicities of 16 of the
remaining 23 systems have been clearly detected in power density
spectra of the ASM light curves; we do not count X0535+262 as a clear
detection in the power spectrum.  The periods of four of the systems
(X0115+634, X0535+262, GX\,304$-$1, and X1145$-$619) are apparent in
the ASM light curves as limited sequences of short transient outbursts
that are separated by the orbital period or small multiples thereof.
The HMXBs X1553-542 and X0331+530 are evident in the ASM light curves
during short transient outbursts, but the ASM yields no evidence of
their orbital periods.  We also find no evidence of the 1.7-d period
of LMC X-3 even though it detectable over most of the 14-year time of
the RXTE mission.

Not unexpectedly, we find that the amplitudes of the variations at the
orbital period, when expressed as fractions of the average intensity,
are typically low for LMXBs and high for HMXBs (see
Tables~\ref{tbl:ampl} and \ref{tbl:uplim}).  This is consistent with
the relatively low rate of detections of orbital modulation in LMXBs
and the high rate in HMXBs.  There are some exceptions such as the
high fractional modulation of the weak LMXB 2S\,0921$-$630 and the low
degree of modulation of the HMXBs LMC X-1 and LMC X-3.  Our results
for 2S\,0921$-$630 suggest that there could be other weak LMXBs with
fractional amplitudes that are not particularly low and that would not
be difficult to detect with an instrument somewhat more sensitive than
the ASM.  Indeed, it is likely that the MAXI instrument on the
International Space Station \citep{maxi09} will produce light curves
useful for extending the search for periodicities in X-ray light
curves and will be able to find many additional periodicities and
orbital periods.

\acknowledgements

We gratefully acknowledge the efforts of the RXTE/ASM science teams at
MIT and NASA/GSFC, and the RXTE mission support groups at GSFC.  We
thank Alex Camacho for computing the upper limits presented in
Table~\ref{tbl:uplim}.  We acknowledge the support of NASA through
Contract NAS 5-30612 (MIT).

\end{document}